\documentclass[referee]{jfm}

\usepackage{amsfonts,amsmath,amssymb,bm,natbib,epstopdf,epsfig,verbatim,
mathtools,mathptmx}
\usepackage{rotating,color}
\usepackage{graphicx,psfrag}

\DeclarePairedDelimiter\floor{\lfloor}{\rfloor}

\ifCUPmtlplainloaded \else
  \checkfont{eurm10}
  \iffontfound
    \IfFileExists{upmath.sty}
      {\typeout{^^JFound AMS Euler Roman fonts on the system,
                   using the 'upmath' package.^^J}%
       \usepackage{upmath}}
      {\typeout{^^JFound AMS Euler Roman fonts on the system, but you
                   dont seem to have the}%
       \typeout{'upmath' package installed. JFM.cls can take advantage
                 of these fonts,^^Jif you use 'upmath' package.^^J}%
      }
  \else
  \fi
\fi

\ifCUPmtlplainloaded \else
  \checkfont{msam10}
  \iffontfound
    \IfFileExists{amssymb.sty}
      {\typeout{^^JFound AMS Symbol fonts on the system, using the
                'amssymb' package.^^J}%
       \usepackage{amssymb}%
       \let\le=\leqslant  \let\leq=\leqslant
         
      }{}
  \fi
\fi

\ifCUPmtlplainloaded \else
  \IfFileExists{amsbsy.sty}
    {\typeout{^^JFound the 'amsbsy' package on the system, using it.^^J}%
     \usepackage{amsbsy}}
    {\providecommand\boldsymbol[1]{\mbox{\boldmath $##1$}}}
\fi

\title[Magnetic confinement of rotating convection]{Confinement of 
rotating convection by a laterally varying magnetic field}
\author[B.~Sreenivasan and V.~Gopinath]%
{Binod  \ls Sreenivasan \ls and \ls
  Venkatesh \ls Gopinath}

\affiliation{Centre for Earth Sciences, Indian Institute of Science, 
Bangalore 560012, India.\\[\affilskip]
}

\pubyear{2016}
\date{}
\begin{document}

\def\apj{\it Astrophys.~J.}
\def\apjs{\it Astrophys.~J.~Suppl.~Ser.}
\def\grl{\it Geophys.~Res.~Lett.}
\def\pre{\it Phys.~Rev.~E}
\def\pre{\it Phys.~Rev.~E}
\def\planss{\it Planet.~Space~Sci.}
\def\nat{\it Nature}

\def\black{\color{black}}
\def\red{\color{red}}
\def\blue{\color{blue}}
\def\Cyan{\color{cyan}}
\def\LimeGreen{\color{green}}

\maketitle

\def\black{\color{black}}
\def\red{\color{red}}
\def\Cyan{\color{cyan}}
\def\LimeGreen{\color{green}}

\begin{abstract}

Spherical shell dynamo models based on rotating convection 
show that the flow within the tangent cylinder is dominated
by an off-axis plume that extends from the inner core boundary to 
high latitudes and drifts
westward. Earlier studies explained the formation of
such a plume in terms of the effect of a uniform axial magnetic field that
 significantly increases the lengthscale of convection
in a rotating plane layer. However, rapidly rotating dynamo
simulations show that the magnetic field within the
tangent cylinder has severe lateral inhomogeneities that
may influence the onset of an isolated plume. Increasing
the rotation rate in our dynamo  simulations (by decreasing the
Ekman number $E$) produces progressively thinner plumes
that appear to seek out the location where the field
is strongest.
Motivated by this result, we examine the linear onset of convection
in a rapidly rotating fluid layer subject to a laterally varying 
axial magnetic
field. A cartesian geometry is chosen where the finite
dimensions $(x,z)$ mimic $(\phi,z)$ in cylindrical
coordinates. The lateral inhomogeneity of the  
field gives rise
to a unique mode of instability where convection is entirely confined
to the peak-field region. The localization of the flow by the magnetic 
field occurs even when 
the field strength (measured by the Elsasser number $\varLambda$) 
is small and viscosity controls the smallest lengthscale of convection. 
The lowest Rayleigh number at which an isolated
plume appears within the tangent cylinder in spherical shell 
dynamo simulations agrees
closely with the viscous-mode Rayleigh number
in the plane layer linear magnetoconvection model. 
The lowest
Elsasser number for plume formation in the simulations is 
significantly higher 
than the onset values in linear magnetoconvection,
which indicates that the viscous--magnetic mode transition point with 
spatially varying fields is displaced
to much higher Elsasser numbers. 

The localized excitation of viscous-mode
convection by a laterally
varying magnetic field provides a mechanism for the formation
of isolated plumes within the Earth's tangent cylinder. The polar
vortices in the Earth's core can therefore be non-axisymmetric.
More generally, this study shows that a spatially varying
magnetic field strongly controls the structure of rotating convection
at a Rayleigh number not much different
 from its non-magnetic value.
\end{abstract}

\section{Introduction}

The Earth's dynamo is powered by thermochemical convection
occurring in its liquid iron outer core. The rapid rotation of the Earth's core
divides convection into two regions, inside and outside the
tangent cylinder. The tangent cylinder is an imaginary cylinder that touches 
the solid inner core and cuts the core surface at approximately latitude
70$^\circ$. The 
tangent cylinder may be approximated by a rotating plane layer in
which convection takes place under a predominantly
axial ($z$) magnetic field and gravity pointing in the downward $z$
direction.  Strongly ageostrophic motions are needed to transport
heat from the inner core boundary to the core--mantle boundary
inside the tangent cylinder \citep{2015jones}, which implies that 
non-magnetic convection inside the tangent cylinder starts
at a Rayleigh number much higher than the threshold value for convection
outside it. At onset, thin upwellings and downwellings aligned
with the axis develop along which
the $z$-vorticity changes sign, in line with the classical
picture of rotating Rayleigh-B\'enard convection in a plane layer.

Observations of secular variation of the Earth's
magnetic field suggest that there are
anticyclonic polar vortices in the core \citep{99olson, 02hulot}. 
Whereas
core flow inversion models support the presence of axisymmetric 
toroidal motions,
it is not clear that relatively small-scale,
non-axisymmetric motions would be dominant
\citep[see][and references
therein]{15holme}.
Non-magnetic laboratory 
experiments that simulate the tangent cylinder
region \citep{03aurnou} show an ensemble of
thin helical plumes extending from the inner core
boundary to high latitudes.
A large-scale anticyclonic zonal flow 
in the polar regions is suggested, the
likely cause of which is a thermal wind \citep{p1987,06sreeni}:
\begin{equation}
2 \Omega \frac{\partial u_\phi}{\partial z}= \frac{g \alpha}{r} 
\frac{\partial T^\prime}{\partial \theta},
\label{thwind1}
\end{equation} 
where $\Omega$ is the angular
velocity about the rotation axis $z$, $g$ is the acceleration due to gravity, 
$\alpha$ is the thermal expansion
coefficient and $T^\prime$
is the temperature perturbation. Equation (\ref{thwind1}) is obtained by
taking the curl of the momentum equation in the inertia-free, inviscid limit.
If the polar regions are slightly warmer than the
equatorial regions due to a build-up of light material, 
(\ref{thwind1}) predicts
an axisymmetric anticyclonic circulation near the poles. It remains
to be seen whether magnetic laboratory experiments
\citep{16aujogue} would support the presence of 
small-scale, non-axisymmetric polar circulation.

Numerical simulations of
the geodynamo \citep[e.g.][]{05sreeni} present a 
different picture from non-magnetic experiments in
that the structure of convection within the 
tangent cylinder is often dominated by an off-axis plume that carries 
warm fluid from the inner core surface to high-latitude regions
(greater than latitude 70$^\circ$). 
This type of convection
also produces a polar vortex because the radially outward flow
at the top of the plume interacts with the background rotation (via the
Coriolis force) to generate a non-axisymmetric, 
anticyclonic flow patch. 
For supercritical convection in the Earth's tangent
cylinder, one or more strong plumes
may be produced which continuously expel magnetic flux from 
high latitudes, 
a process that may be inferred from observation of the rather weak flux 
in this region \citep{2000jackson} or the location of
the persistent magnetic flux patches just outside the tangent cylinder 
\citep{07gubetal}. To understand the physical origin of the
isolated plumes within the tangent cylinder, \S 2
focuses on their \emph{onset}; that is, the regime of their
first appearance.

The linear theory of magnetoconvection \citep{61chandra}
predicts that onset in a rotating plane layer occurs
either as thin viscously controlled columns or 
large-scale magnetic rolls (see, for example,
the structures in figure \ref{fig5}(\emph{b}) 
and (\emph{d}) in \S 3).
\cite{06sreeni} equate the critical Rayleigh numbers for 
the viscous and magnetic
branches of onset to obtain the transition point Elsasser number 
$\varLambda \approx 7.2 E^{1/3}$, where $E$ is the Ekman number. 
(Here, $\varLambda$ measures the uniform magnetic field strength
 and $E$ is the ratio of viscous to Coriolis forces).
If the momentum diffusivity is given a \lq turbulent' value of the order of
the magnetic diffusivity, then $E \sim 10^{-9}$, so
the viscous--magnetic cross-over value is 
$\varLambda \approx 7.2 \times 10^{-3}$.
As this is much less than the observed dipole field at the 
Earth's core--mantle
boundary, \cite{06sreeni} propose that the off-axis
plumes within the tangent cylinder may be in the 
large-scale magnetic
mode. However, these arguments rely on the assumption of a uniform axial
magnetic field permeating the fluid layer, 
whereas rapidly rotating dynamo simulations show that the
magnetic field has severe axial and lateral inhomogeneities. 
An important aim of our study is to see whether isolated
plumes can form via
confinement of viscous-mode convection by the naturally
occurring, \emph{laterally varying} magnetic 
field distribution within the tangent cylinder. This
necessitates a comparative study across $E$
of plume onset in dynamo simulations
(\S \ref{dynamo}).

The onset of convection in three-dimensional physical systems
has been well understood from one-dimensional linear onset theory. 
Early experiments on the onset of convection in a 
rotating cylinder containing mercury heated from below
and placed in a uniform axial magnetic field 
\citep{57nakagawa} show that the measured critical Rayleigh number
agrees closely with that predicted by one-dimensional plane 
layer onset theory \citep{61chandra}.
Subsequently, MHD instabilities have been extensively
studied
using spatially varying imposed fields of the form
${\bm B}=B_0 s \, \hat{\bm{\phi}}$ in cylindrical
coordinates ($s,\phi,z$)
\citep{67malkus,79soward,03jonesetal} or more complex 
fields thought to be relevant to rotating dynamos
 \citep{83fearn,90kuang,95zhang,95longbottom,97tucker,11sreeni}. 
In these studies, the back-reaction 
of the mean field on convection via the linearized Lorentz
force is the main point of interest, while the generation of the mean field
itself is decoupled from this process. Although an incomplete
representation of the nonlinear dynamo, linear magnetoconvection provides
crucial insights into how the field changes the 
structure of the flow at onset. For a field that is either uniform or of
a lengthscale comparable to the depth of the fluid layer, 
large-scale magnetically controlled convection
sets in at small Elsasser numbers $\varLambda=O(E^{1/3})$
\citep[e.g.][]{95zhang,03jonesetal}. On the other hand, 
if the lengthscale
of the field is small compared to the layer depth 
as rapidly rotating dynamo models suggest,
the viscous--magnetic mode transition point is displaced to Elsasser numbers 
 $\varLambda=O(1)$ or higher \citep{15sreeni}. The fact that small-scale
convection is possible for a wide range of $\varLambda$ suggests that
convection in the Earth's core may operate in the viscous mode.  

Linear stability models that consider
variation of the basic state variables along
two coordinate axes
\citep{11theo} resolve the perturbations in  
two finite dimensions,
while the third dimension is of infinite extent.
Recent examples of
linear onset models where perturbations are resolved in more 
than one direction
include that of double-diffusive convection in a rectangular
duct with or without a longitudinal flow \citep{12benhadid}, 
and quasi-geostrophic
convection in a cylindrical annulus with the gravity pointing radially
outward \citep{13calkins}. Models of rotating convection
subject to laterally varying magnetic fields are not available.
Motivated by the \emph{onset} of localized convection
within the tangent cylinder in nonlinear dynamos, \S 3 examines onset 
 in a rotating plane layer subject to a laterally
varying magnetic field. The finite vertical ($z$) dimension and one
horizontal ($x$) dimension in cartesian coordinates 
mimic the axial ($z$) and azimuthal ($\phi$) dimensions
respectively in cylindrical coordinates.

For the classical case of convection under a uniform 
field of $\varLambda=O(1)$,
the scale of convection perpendicular to the rotation axis $L_\perp$ is
significantly increased, and this reduces the Ohmic and viscous
dissipation rates. As the work done by the buoyancy force need not
be high in order to maintain convection, the critical Rayleigh number $Ra_c$
is much lower than for non-magnetic convection \citep{kr2002}.
On the other hand, if convection under a spatially inhomogeneous field
is viscously controlled so that $L_\perp$ is much smaller than the axial
lengthscale of columns, $Ra_c$ would be comparable to its
non-magnetic value. The lengthscale of convection thus has 
implications for the power
requirement of a rotating dynamo. 
An obvious counterpoint to this argument is that of
subcritical behaviour, wherein saturated (strong-field)
numerical dynamos survive at a Rayleigh number lower than that
required for a seed field to grow \citep[e.g.][]{08kuang,11sreeni,13hori}.
The role of the self-generated magnetic field in lowering the threshold 
for convection
appears to be consistent with the classical theory of
convective onset under a uniform magnetic field \citep{61chandra} that 
predicts a significant decrease in critical Rayleigh
number from its non-magnetic value. 
Numerical dynamo simulations at $E \sim 10^{-4}$, however, show
that subcritical behaviour is
preferred for relatively small magnetic Prandtl numbers $Pm \le 1$
rather than for $Pm>1$ \citep{09morin,11sreeni}, which indicates
that a relatively large ratio of the 
inertial to Coriolis forces in the equation of motion (measured by the
Rossby number $Ro=EPm^{-1} Rm$, where $Rm$ is the magnetic
Reynolds number) may promote subcriticality.
Furthermore, \cite{11sreeni} show  that the depth of subcriticality $d_{sub}$
in rotating spherical dynamos is strongly influenced by
the kinematic boundary condition. No-slip boundaries produce
dominant columnar convection via Ekman pumping, but give a $d_{sub}$ value
 that is much smaller
than for stress-free boundaries where large-scale zonal flows
dominate even in slightly supercritical convection. 
Dynamo calculations
at lower $E$ would help ascertain  whether $d_{sub}$ remains
relatively constant or decreases with decreasing Ekman number.
Another point of relevance here is that the back-reaction of
the magnetic field on the columnar flow need not 
drastically change 
the transverse lengthscale of convection $L_\perp$. 
\cite{14sreeni} find that
the magnetic field enhances the relative
kinetic helicity between
cyclones and anticyclones, a process that is 
essentially independent of $L_\perp$. 
Indeed, saturated spherical dynamo models
show that the magnetic field does not appreciably
increase $L_\perp$ from its non-magnetic value \citep[see,
for example,][]{15sreeni}. 
In short, the magnetic
field can enhance helical fluid motion while
preserving the small-scale structure produced by rapid rotation.

Present-day dynamo models mostly operate in parameter regimes where the viscous
and Ohmic dissipation rates are comparable in magnitude. 
If Ohmic dissipation at
small lengthscales $L_\perp$ must dominate over 
viscous dissipation as in liquid metal magnetohydrodynamic
turbulence \citep{01davidson}, the magnetic diffusivity $\eta$
must far exceed the momentum diffusivity $\nu$, so that
$Pm=\nu/\eta<<1$. Dynamos operating
in this regime are very likely turbulent, with a well-defined
energy cascade from the energy injection scale to the Ohmic
dissipation scale. Geodynamo models typically operate at $Pm \sim 1$
\citep[e.g.][]{07chrwicht} where the turbulent value of $\nu$ is assumed to
match $\eta$. Low-$E$, low-$Pm$ models are rare
because of the computational effort involved in solving them,
but linear magnetoconvection models with spatially varying fields
are possible at these parameters. Apart from predicting
whether convection in the Earth's core operates in small scales, these
models also give the peak local Elsasser numbers in the core that
would still yield a volume-averaged Elsasser number $\overline{B^2}$
of order unity. The analysis of rapidly 
rotating convection under a spatially varying 
magnetic field is partly motivated by these ideas. 
 
In this study, it is shown that a laterally inhomogeneous
magnetic field gives rise to isolated columnar vortices in a
rotating plane layer at the onset of convection. 
This mode of onset is linked to the formation
of isolated plumes within the tangent cylinder 
in convection-driven 
dynamos. \S \ref{dynamo} presents 
nonlinear dynamo simulations 
where strongly localized convection appears
within the tangent cylinder. Since
the critical Rayleigh number is much higher within the tangent cylinder
than outside it, supercritical dynamo simulations present the opportunity
to visualize the onset of isolated plumes within the tangent cylinder.
\S \ref{magneto} considers the linear onset of convection
 in a rotating plane layer of finite aspect ratio subject to a laterally
varying axial magnetic field. The onset of localized
convection within the tangent cylinder is then interpreted in the
light of the linear magnetoconvection results. 
The main results of this paper are summarized in \S \ref{concl}.

\section{Nonlinear dynamo simulations}
\label{dynamo}
The aim of the spherical shell dynamo simulations 
is to obtain the regime for 
onset of localized convection within
the tangent cylinder. 
In the Boussinesq approximation \citep{kr2002}, we 
consider the dynamics an electrically conducting fluid 
confined between two concentric, co-rotating spherical surfaces
whose radius ratio is 0.35.
The main body forces acting
 on the fluid  are the thermal buoyancy force, 
the Coriolis force originating from the
background rotation of the system and the Lorentz 
force arising from the interaction between
the induced electric currents and the magnetic fields.
The non-dimensional magnetohydrodynamic (MHD) equations 
for the velocity ${\bm u}$, magnetic field
$ \bm{B}$ and temperature $T$ are
\begin{align}
E Pm^{-1}  \Bigl(\frac{\partial {\bm u}}{\partial t} + 
(\nabla \times {\bm u}) \times {\bm u}
\Bigr)+  {\bm \hat{z}} \times {\bm u} &= - \nabla p^\star +
Ra \, Pm Pr^{-1} \, T \, {\bm r} \,  \nonumber\\ 
& \quad +  (\nabla \times {\bm B})
\times {\bm B} + E\nabla^2 {\bm u}, \label{momentum} \\
\frac{\partial {\bm B}}{\partial t} &= \nabla \times ({\bm u} \times {\bm B}) 
+ \nabla^2 {\bm B},  \label{induction}\\
\frac{\partial T}{\partial t} +({\bm u} \cdot \nabla) T &=  Pm Pr^{-1} \,
\nabla^2 T,  \label{heat1}\\
\nabla \cdot {\bm u}  &=  \nabla \cdot {\bm B} = 0.  \label{div}
\end{align}
The modified pressure $p^\star$ in equation (\ref{momentum}) 
is given by $p+\frac1{2} {E}{Pm^{-1}} |{\bm u}|^2$,
where $p$ is the fluid pressure. The velocity  satisfies 
the no-slip condition at the boundaries and the
magnetic field matches a potential field at the outer boundary.
Convection is set up in the shell
by imposing a temperature difference
between the boundaries. The basic state temperature
distribution is given by $T_0(r)=\beta/r$, where
$\beta=r_i r_o$. Equations (\ref{momentum})--(\ref{div})
are solved by a dynamo code that uses spherical
harmonic expansions in ($\theta,\phi$) and finite
difference discretization in $r$ \citep{07willis}.
The radial grid points are located at the zeros of a
Chebyshev polynomial and are clustered near the boundaries.
 
The dimensionless parameters in equations 
(\ref{momentum})--(\ref{heat1}) are the Ekman number $E$, 
the modified Rayleigh number $Ra$, Elsasser number $\varLambda$,
Prandtl number $Pr$ and magnetic Prandtl number $Pm$, 
which are defined as follows:
\begin{equation}
 E = \frac{\nu}{2 \varOmega L^2}, \, 
Ra = \frac{g \alpha \Delta T L}{2 \varOmega \kappa}, 
\, Pr= \frac{\nu}{\kappa}, \, Pm= \frac{\nu}{\eta},
\label{parameters}
\end{equation}
where $L$ is the spherical shell thickness, $\nu$ is the 
kinematic viscosity, $\rho$ is the density, 
$\kappa$ is the thermal diffusivity, $\eta$
is the magnetic diffusivity,
$g$ is the gravitational acceleration, $\alpha$
is the coefficient of thermal expansion, $\Delta T$ is the
superadiabatic temperature difference between the boundaries, 
$\varOmega$ is the angular velocity of background rotation and
$\mu_0$ is the magnetic permeability. 
The ratio $Pm Pr^{-1}$ is also called the Roberts number, $q$.
The Elsasser number $\varLambda= B^2/2 \varOmega \rho \mu_0 \eta$
is an output that measures the volume-averaged 
strength of the self-generated 
magnetic field in the model. In addition,
the Elsasser number $\varLambda_z$ based on the measured peak
axial ($z$) magnetic field within the tangent cylinder
is also defined.

Two parameter regimes are considered in this study:
(\emph{a}) $E= 5\times 10^{-5}$, $Pr=Pm=5$, and (\emph{b}) 
$E=5 \times 10^{-6}$, $Pr=Pm=1$.
The Roberts number $q=1$ in both regimes, but at 
the higher $E$ the choice of the larger $Pr=Pm$ 
keeps nonlinear inertia small in the simulation \citep{06sreejon}. 
Runs for $E=5 \times 10^{-5}$ are done with 96 finite difference
grid points in radius and a maximum 
spherical harmonic degree $l=72$. For $E=5 \times 10^{-6}$, 192
radial grid points and a spectral cut-off of $l=160$
are used. Simulations in both parameter regimes 
produce strongly dipole-dominated
magnetic fields.

\begin{figure}
\begin{center}
\resizebox{14 cm}{!} 
{\includegraphics{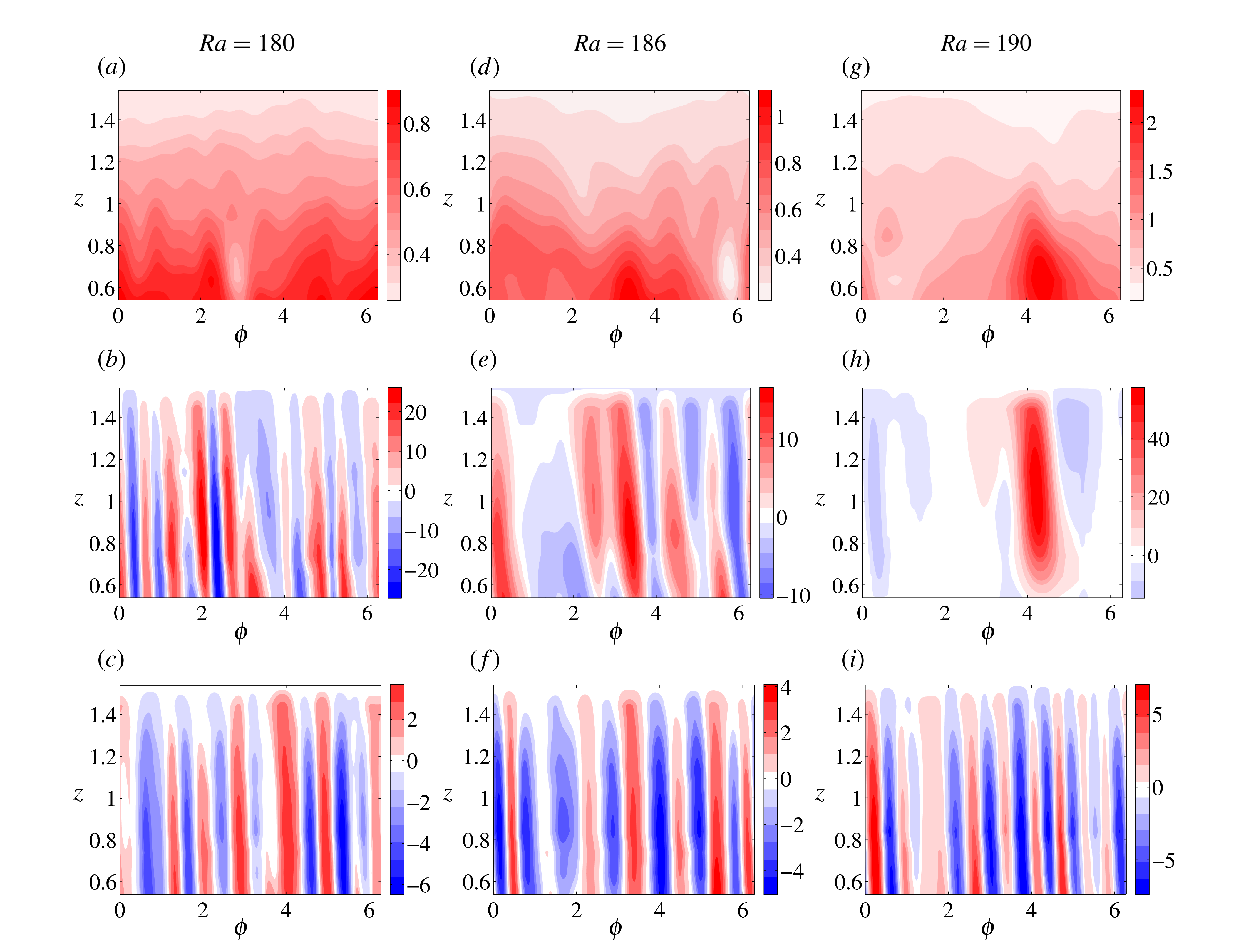}}\\
\caption{Cylindrical section ($z-\phi$) plots within the tangent cylinder
 of the $z$-magnetic field (top panels),
 dynamo $z$-velocity (centre panels) and
 non-magnetic $z$-velocity (bottom panels),
for $E=5 \times 10^{-5}$, $Pr=Pm=5$ and three Rayleigh numbers $Ra$ near
onset of magnetic convection.
No-slip, electrically insulating boundaries are used. The plots are
shown at cylinder radii $s=0.33$ (\emph{a,b}), $s=0.18$ (\emph{d,e}),
$s=0.21$ (\emph{g,h}) and $s=0.3$ (\emph{c,f,i}).}
\label{fig1}
\end{center}
\end{figure}

 The focus of attention in this study is on the onset of 
localized convection within the tangent cylinder. For
each dynamo calculation, an equivalent non-magnetic calculation
is done for which only the
momentum and temperature equations are stepped forward in time.
For $E=5 \times 10^{-5}$, convection starts 
in the tangent cylinder at $Ra \approx 140$ in both
dynamo and non-magnetic runs, which indicates that the magnetic
field does not alter the critical Rayleigh
number for onset. (Convection outside the tangent cylinder
sets in at a much lower value of $Ra=29.61$). 
At $Ra=180$, the tangent cylinder is filled with 
upwellings and downwellings, 
although the effect
of the magnetic field is visible in the enhanced velocity
in plumes (compare figures \ref{fig1}\emph{b} and \emph{c}). 
The $z$-magnetic field 
appears to have mostly diffused in from outside, 
where dynamo action via columnar convection occurs at much lower
Rayleigh number (figure \ref{fig1}\emph{a}). Close to onset, 
the magnetic field $B_z$ is not affected much
by the plumes, which is why this diffused 
field is largely homogeneous in
the azimuthal ($\phi$) direction. At $Ra=186$, 
convection is strong enough to cause some
lateral inhomogeneity in the magnetic field. Patches of $B_z$ form at the
base of the convection zone (figure \ref{fig1}\emph{d})
 because of convergent flow at the base of plumes. 
Dominant upwellings (in red) form over the flux patches 
while weak convection
exists in other areas (figure \ref{fig1}\emph{e}). 
At $Ra=190$, the highly inhomogeneous
field patch that develops at the bottom concentrates 
convection over it and
wipes out convection in the rest of the fluid layer 
(figure \ref{fig1}\emph{g,h}). 
A progressive
enhancement of $B_z$ occurs until a threshold field strength is reached,
upon which convection is supported only in the strong-field region. 
At subsequent
times, the flow follows the path of the peak magnetic field.
The non-magnetic runs
at $Ra=186$ and $Ra=190$ show a uniformly distributed 
axial flow structure 
(figure \ref{fig1}\emph{f,i}), which suggests 
that the confinement of
 convection in the dynamo is due to the laterally inhomogeneous 
magnetic field that forms within the tangent cylinder. 

\begin{figure}
\begin{center}
\resizebox{14 cm}{!} 
{\includegraphics{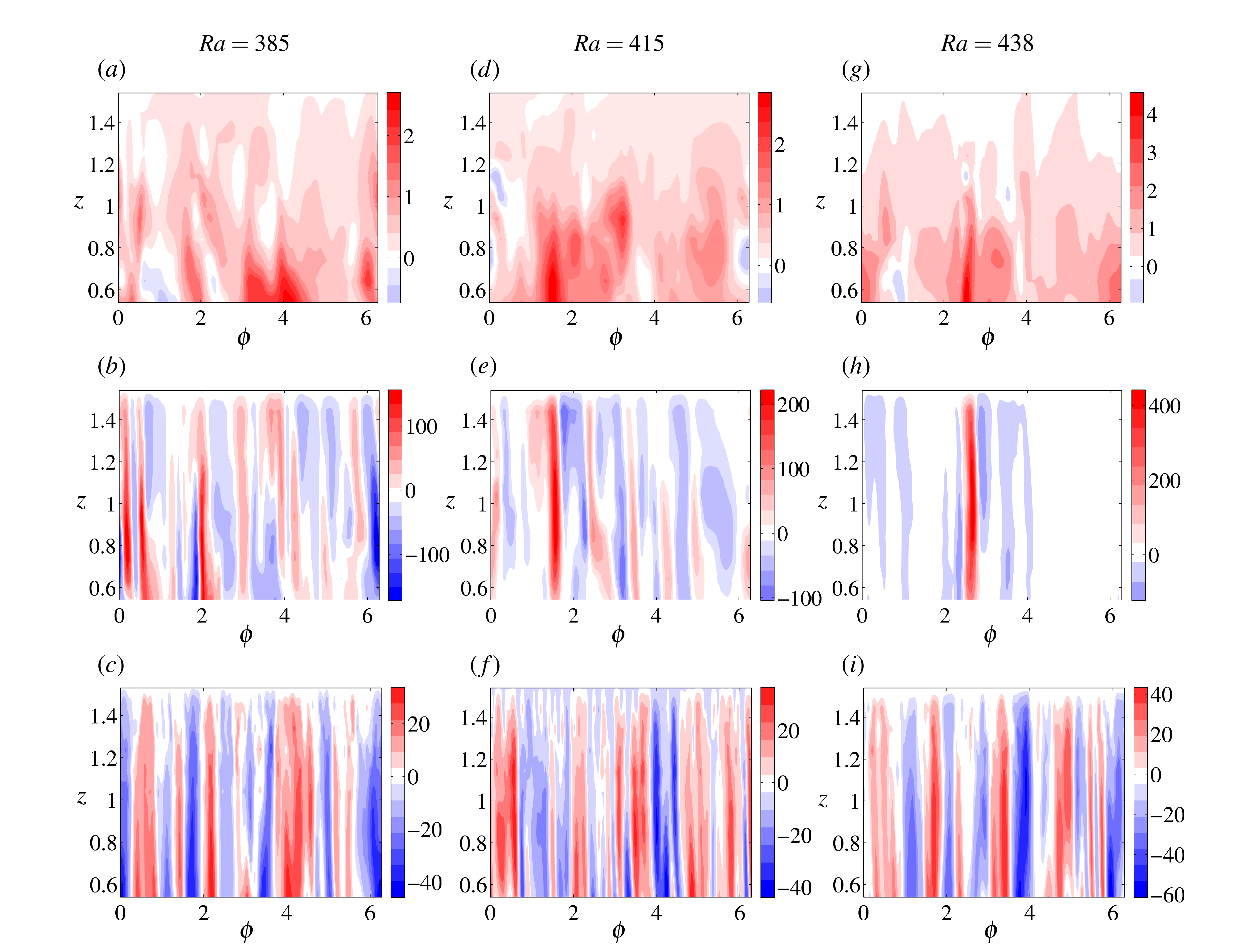}}\\
\caption{Cylindrical section ($z-\phi$) plots within the tangent cylinder
 of the $z$-magnetic field (top panels),
dynamo $z$-velocity (centre panels) and non-magnetic $z$-velocity (bottom panels),
for $E=5 \times 10^{-6}$, $Pr=Pm=1$ and three Rayleigh numbers $Ra$ near
onset of magnetic convection.
No-slip, electrically insulating boundaries are used. The plots are
shown at cylinder radii $s=0.35$ (\emph{a,b}), $s=0.31$ (\emph{d,e}),
$s=0.33$ (\emph{g,h}) and $s=0.3$ (\emph{c,f,i}).}
\label{fig2}
\end{center}
\end{figure}

For $E=5 \times 10^{-6}$, convection outside the 
tangent cylinder starts at $Ra=50.18$.
Figure \ref{fig2}(\emph{b}) shows that at $Ra=385$,
small-scale convection is uniformly 
distributed inside the tangent cylinder.
A comparison of the $z$-velocities in the dynamo and non-magnetic
runs (figure \ref{fig2}\emph{b,c}) shows that the magnetic field intensifies
the flow even as its small-scale structure 
is preserved.  
The scale of the lateral variation of $B_z$ seen in figure \ref{fig2}(\emph{d})
($Ra=415$) is fixed by the pre-existing small-scale velocity field interacting
with the field diffusing from outside the tangent cylinder, and
this can  explain why the transverse lengthscale of $B_z$ is 
appreciably smaller compared to that at the higher Ekman number. 
The small-scale patches of $B_z$ in turn concentrate
small-scale $u_z$ over them, although convection is still active in 
other regions. The formation of isolated plumes causes a skewness in 
$u_z$, with the peak upwelling velocity being approximately 
twice the downwelling velocity 
(figure \ref{fig2}\emph{e}). As $Ra$ is increased to 438, $B_z$ is strong
enough to concentrate a small-scale plume over it and
suppress convection elsewhere (figure \ref{fig2}\emph{g,h}).
The marked decrease in the azimuthal lengthscale of the plume with
decreasing Ekman number 
(figures \ref{fig1}\emph{h} and \ref{fig2}\emph{h}) suggests  
 that, while the plume is magnetically confined, its width 
(lengthscale perpendicular to the rotation axis, $L_\perp$) may be
controlled by the fluid viscosity. 
As with the higher Ekman number, the non-magnetic simulations
retain the uniformly distributed axial flow structure from the onset
of convection (figure \ref{fig2}\emph{c,f,i}). 

\begin{figure}
\begin{center}
\resizebox{14 cm}{!} 
{\includegraphics{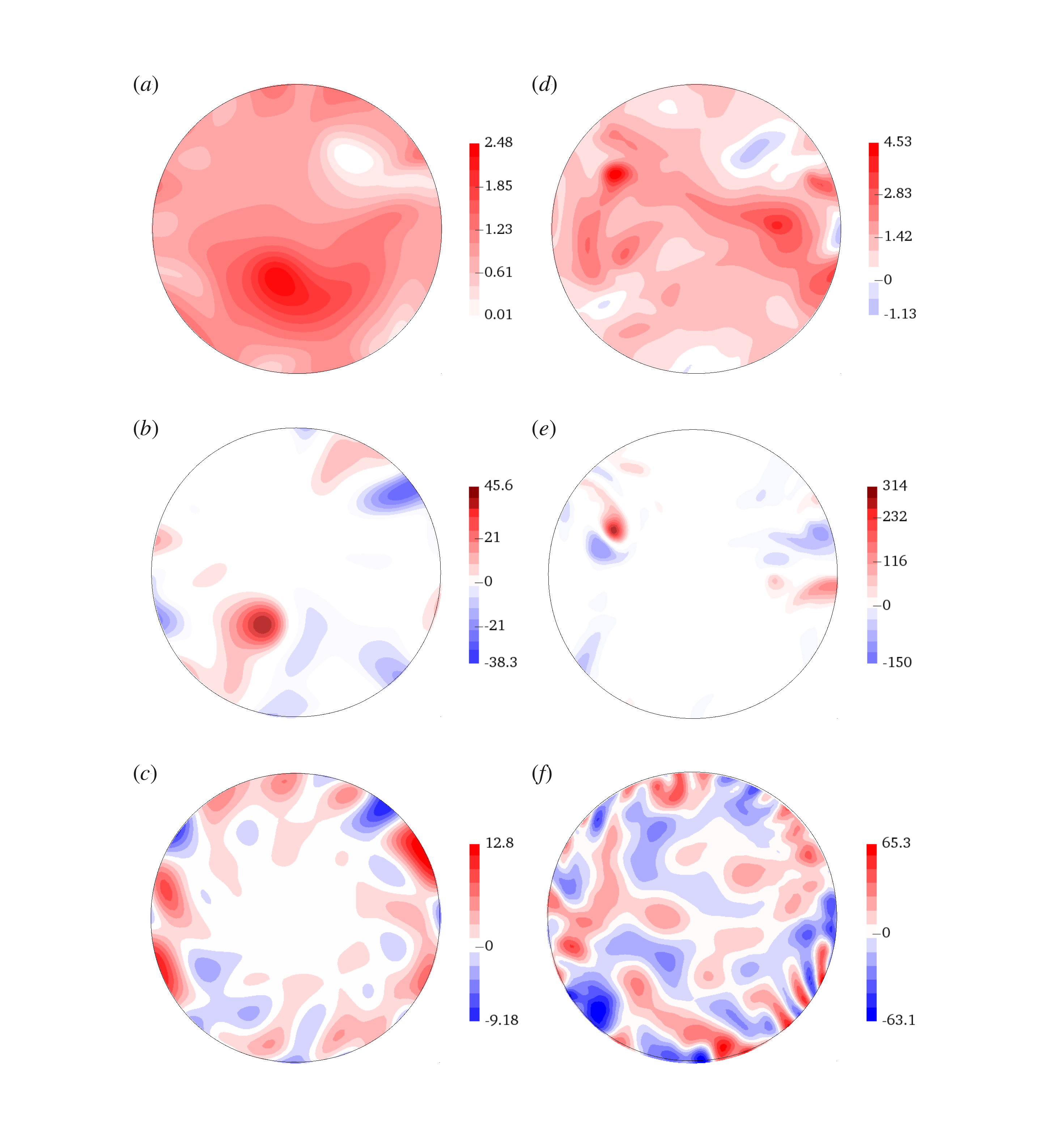}}\\
\caption{Horizontal ($z$) section plots within
 the tangent cylinder of 
the axial magnetic field $B_z$
at $z=0.9$ (top panel), the axial velocity $u_z$ for the dynamo 
at $z=1.4$ (centre panel) and
$u_z$ for non-magnetic convection at $z=1.4$ (bottom panel). The
periphery of these sections are at colatitude $21.04^\circ$ ($z=1.4$)
and $30.9^\circ$ ($z=0.9$). The cylindrical 
radius in all plots is in the
range $[0,0.5384]$ in dimensionless units.
(\emph{a-c}) $E=5 \times 10^{-5}$, $Pr=Pm=5$, $Ra=190$; 
(\emph{d-f}) $E=5 \times 10^{-6}$, $Pr=Pm=1$, $Ra=438$.
No-slip, electrically insulating boundary conditions are used.}
\label{fig3}
\end{center}
\end{figure}

Figure \ref{fig3} shows horizontal ($z$) section plots 
within the tangent cylinder of $B_z$ and $u_z$ for the
two Ekman numbers at onset of the off-axis plume. The non-magnetic
$u_z$ is provided for comparison. As $B_z$ is concentrated
near the base of the convection zone, the strong correlation 
between the magnetic field and the flow is clearly
visible by looking at two different sections, $z=0.9$
for $B_z$ and $z=1.4$ for $u_z$. The decrease in
plume width $L_\perp$ at the lower Ekman number is evident by 
comparing the  section plots
of $u_z$ at the same $z$ (figure \ref{fig3}\emph{b,e}). 
It is plausible that the magnetic field 
locally reduces the Rayleigh
number for convection from its non-magnetic value, 
upon which the plume finds the location where
the field is strongest. A strongly
supercritical ($Ra=350$) dynamo simulation suggests that this idea
deserves consideration, as the dominant upwelling in the
tangent cylinder continuously migrates to the location of the peak 
magnetic field in a period of less than $\sim$ 0.1 magnetic
diffusion time. Further studies at higher $Ra$ are necessary to obtain
the regime where the $B_z$--$u_z$ correlation within the tangent cylinder 
completely breaks down.

Table \ref{table0} presents the parameters and some
key properties of the dynamo simulations performed
for the two Ekman numbers. The volume-averaged Elsasser number $\overline{B^2}$ 
($ \sim 1$ in all runs)
does not give any insight into the onset of the 
localized plume in the tangent cylinder;
on the other hand, the Elsasser number $\varLambda_z$ 
calculated
based on the peak $B_z$ value in the tangent cylinder 
shows a clear increase
at plume onset. The field components $B_s$
and $B_\phi$ are a factor $\approx 3$ lower than $B_z$. 

\begin{table}
	\centering
	\begin{tabular}{ c c c c c }
      $E$	 & $Ra$ & $Rm$ & $\overline{B^2}$ & $\varLambda_z$ \\ 
\hline
 & 180 & 66.19 & 0.9 & 1.34 \\
$E=5 \times 10^{-5}$ & 186 & 68.11 & 0.92 & 2.94 \\
 & 190 & 68.58 & 0.97 & 3.49 \\
\hline
 & 385 & 164.95 & 1.25 & 6.76 \\
$E=5 \times 10^{-6}$ & 415 & 181.81 & 1.39 & 7.92 \\
 & 438 & 186.05 & 1.46 & 16.81 \\
\hline
\end{tabular}
\caption{Summary of the dynamo calculations for two Ekman numbers ($E$)
at $q=1$ with
no-slip, isothermal and electrically insulating boundary conditions.
Here $Ra$ is the modified Rayleigh number, 
$Rm$ is the magnetic Reynolds number obtained from the root mean
square (rms) value of the velocity and $\varLambda_z$ is the
Elsasser number given by the square of the peak value of $B_z$
in the tangent cylinder. }
\label{table0}
\end{table}

A key issue that arises from the nonlinear dynamo simulations is
whether the isolated plumes that form within the tangent cylinder 
are viscously or magnetically controlled. Although
it may appear from the simulations at $E=5 \times 10^{-5}$
that the magnetic field increases the scale of 
convection at plume onset (see figure \ref{fig1}\emph{b} and \emph{h}), 
a comparison across Ekman numbers shows
that the plume width decreases with decreasing Ekman number
(figure \ref{fig3}\emph{b,e}).
In addition, the Rayleigh number for plume onset
within the tangent cylinder increases with
decreasing Ekman number. These findings suggest that
the onset of isolated plumes within the tangent cylinder
 is controlled by the fluid viscosity. 

As the sloping boundaries (top and bottom caps) 
of the tangent cylinder themselves prevent
perfect geostrophy, the critical Rayleigh number $Ra_c$
and wavenumber $k_c$ at
which non-magnetic convection sets in may
 not be faithfully reproduced by a plane layer
linear onset model. On the other hand, if a laterally varying magnetic field
strongly localizes convection in the tangent cylinder, 
a plane layer
magnetoconvection model could be a good approximation for the onset
of magnetic convection in the tangent cylinder because 
the change of boundary curvature across a thin plume is small. 
Therefore, a study of convective onset in a plane layer under a 
laterally varying magnetic field is justified. This study is presented
in the following section.

\section{Linear magnetoconvection model}
\label{magneto}

\subsection{Problem set-up and governing equations}
\label{setup}
\begin{figure}
\begin{center}
\resizebox{14 cm}{!} {\includegraphics{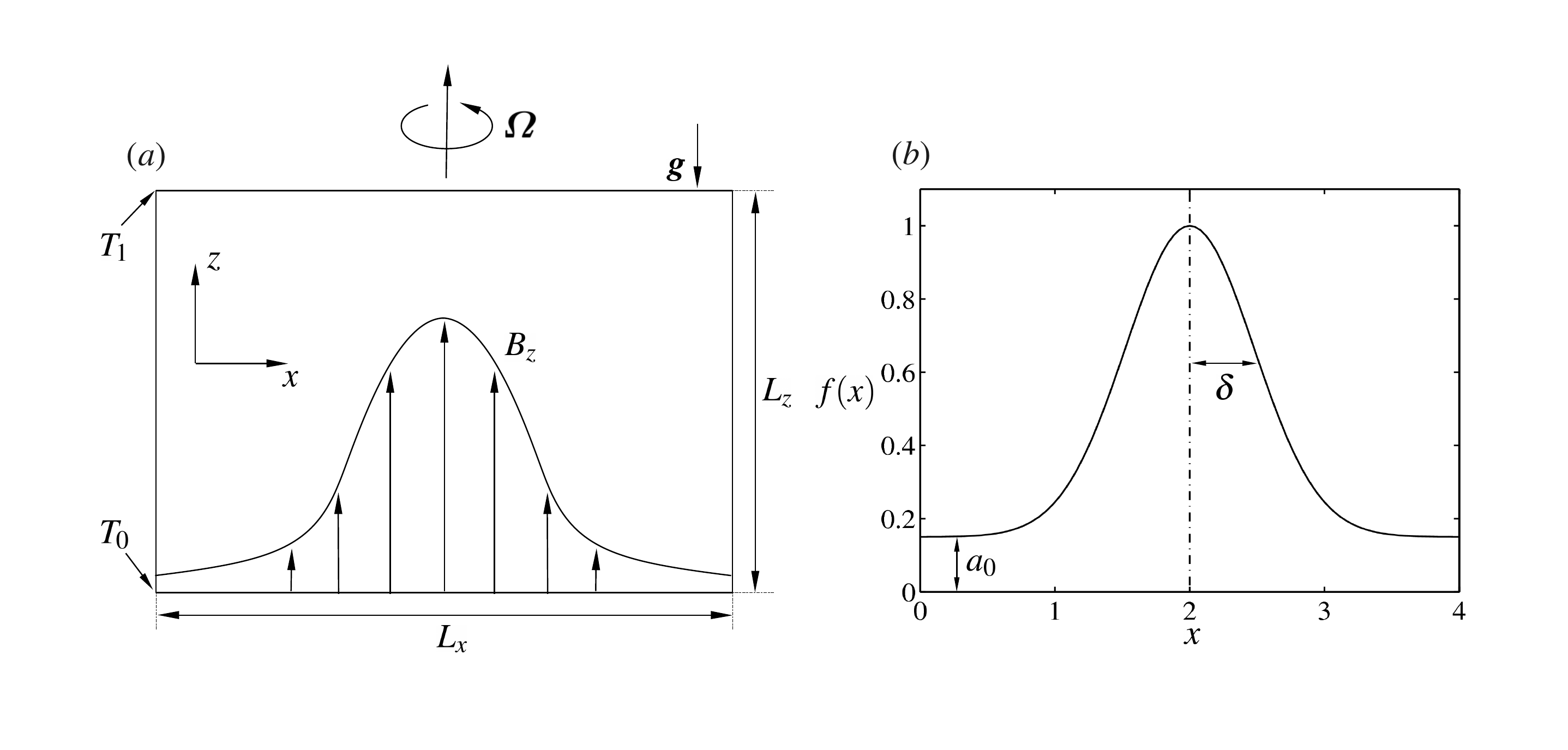}}\\
\caption{(a) Schematic of plane layer rotating magnetoconvection produced 
by a constant adverse temperature gradient
under a laterally varying axial magnetic field. (b) The profile of 
$B_0$ in the layer from equation (\ref{bprofile}), with $a_0=0.15$,
$a_1=0.85$, $c=2$ and $\delta=0.48$.}
\label{fig4}
\end{center}
\end{figure}

We consider an electrically conducting fluid in 
a plane layer of finite aspect ratio, where the vertical ($z$) and a
horizontal ($x$) lengthscale are known and  
the third direction ($y$) is of infinite horizontal extent. The $z$ and
$x$-directions mimic the axial ($z$) and azimuthal ($\phi$) directions
in cylindrical polar coordinates $(s,\phi,z)$. The basic state
temperature gradient across the layer sets up convection under gravity 
${\bm g}$
that acts in the negative $z$ (downward) direction. The system rotates 
about the $z$-axis. The fluid layer is permeated by a laterally
varying magnetic field of the form
\begin{equation}
{\boldsymbol B}_0=\mathcal{B}_0 f(x) \hat {\boldsymbol{z}}; \,\,\,
f(x)= a_0+ a_1 \exp \bigl[-(x-c)^2/2\delta^2\bigr],
\label{bprofile}
\end{equation}
where $\mathcal{B}_0$ is a reference magnetic field strength, 
$a_0$, $a_1$ and $c$ are constants and $\delta$ is the horizontal lengthscale 
of the magnetic field. The problem set-up is shown in figure \ref{fig4}.

In the Boussinesq approximation, the following 
linearized MHD
equations govern the system:
\begin{align}
E Pm^{-1} \frac{\partial \bm{u}}{\partial t} + 
\hat{\bm{z}} \times \bm{u} = -{\bm \nabla} p +
\varLambda\bigl[({\bm \nabla} \times \bm{B}_0) \times \bm{b} + 
({\bm \nabla} \times \bm{b})
\times \bm{B}_0 \bigr] \nonumber \\
+ Pm Pr^{-1} Ra \theta \hat{\bm{z}} + E  \nabla^2 \bm{u}, \label{mom1}\\
\frac{\partial \bm{b}}{\partial t} = {\bm \nabla} \times (\bm{u} \times \bm{B}_0) 
+ \nabla^2 \bm{b}, 
\label{ind1}\\
\frac{\partial \theta}{\partial t} = \bm{u} {\bm \cdot} \hat{\bm{z}} 
+ Pm Pr^{-1} \nabla^2 \theta,
\label{temp1}\\
{\bm \nabla} {\bm \cdot} \bm{u} = \bm{\nabla} {\bm \cdot} \bm{b}=0, 
\label{ubcont} 
\end{align}
The dimensionless parameters $E$, $Ra$, $Pm$ and $Pr$ in equations 
(\ref{mom1})--(\ref{temp1}) have the same definitions as in 
(\ref{parameters}), except that the spherical shell thickness $L$ is
replaced by the plane layer depth $L_z$. The Elsasser
number $\varLambda= \mathcal{B}_0^2/2 \varOmega \rho \mu_0 \eta$
is defined based on the reference magnetic field strength.

By applying the operators $({\bm \nabla} \times)$ and 
$({\bm \nabla} \times {\bm \nabla} \times)$ to the momentum equation 
(\ref{mom1})
and $({\bm \nabla} \times)$ to the induction equation (\ref{ind1}) and taking the 
$z$-components of the equations,
the behaviour of the five perturbation variables -- velocity, vorticity, 
magnetic field, electric current
density and temperature -- can be obtained. As the Roberts number
$q$ is set to unity throughout
this study, the onset of convection with an axial magnetic field is expected
to be stationary for a wide range of Ekman numbers \citep{15kelig}.
Furthermore, this study aims to investigate the
structure of convection at onset and seek comparisons with the 
long-time convection pattern within the tangent cylinder in saturated
(quasi-steady) nonlinear dynamos. 
The time dependence
of the perturbations is therefore not considered, and  
solutions are sought in the following form:
\begin{eqnarray}
\bigl[u_z',\omega_z',b_z', j_z',\theta' \bigr](x,y,z) = \bigl[u_z(x,z), 
\omega_z(x,z), b_z(x,z),
\nonumber\\
j_z(x,z), \theta(x,z) \bigr] \exp \, (iky),
\end{eqnarray}
where $k$ is the wave number in the $y$-direction. 
After introducing this solution into the governing
equations, the following 
system of differential equations is obtained:
\begin{eqnarray}
E \bigl(D_x^2+D_z^2-k^2\bigr)^2 u_z + \Lambda \bigl[f(x)\bigl(D_x^2+D_z^2- 
k^2\bigr)D_z b_z    \nonumber \\ 
+ f''(x) D_z b_z + 2 f''(x) D_x b_x+ f'(x) D_x^2 b_x 
 + 2 f'(x) D_{xz}^2 b_z  \nonumber \\ 
 + f'''(x) b_x - k^2 f'(x) b_x- f'(x) D_z^2 b_x   \bigr] - D_z \omega_z - 
q Ra \bigl(D_x^2-k^2 \bigr) \theta &=& 0, \label{mom2}\\
D_z u_z + \Lambda \bigl(f(x) D_z j_z +  f'(x) D_z b_y  \bigr) + 
E \bigl(D_x^2+D_z^2-k^2\bigr) \omega_z &=& 0, \label{vort2}\\
q \bigl(D_x^2 + D_z^2-k^2\bigr) \theta +u_z &=& 0, \label{temp2}\\
f(x) D_z u_z - f'(x) u_x + \bigl(D_x^2+D_z^2-k^2\bigr) b_z &=& 0, \label{ind2}\\
f(x) D_z \omega_z + f'(x) D_z u_y + \bigl(D_x^2+D_z^2-k^2\bigr) j_z &=& 0, 
\label{current2}
\end{eqnarray}
where $D_x=\partial/\partial x$ and $D_z=\partial/\partial z$. The variables
$u_x$, $u_y$, $b_x$, $b_y$ are related to the eigenfunctions $u_z$, $\omega_z$,
$b_z$, $j_z$ by the identities 
\begin{eqnarray}
-[\nabla_{H}^2] u_x &=& D_x D_z u_z + i k \omega_z, \, -[\nabla_{H}^2] u_y = i k D_z u_z - D_x \omega_z,\\
-[\nabla_{H}^2] b_x &=& D_x D_z b_z + i k j_z, \,\,\, -[\nabla_{H}^2] b_y = i k D_z b_z - D_x j_z,
\end{eqnarray}
where $\nabla_{H}^2=D_x^2-k^2$ is the horizontal Laplacian.

The stability calculations are performed with both stress-free and no-slip
boundaries on $z$. Electromagnetic conditions are insulating at the top
and bottom, although one set of calculations with mixed (bottom
perfectly conducting and top insulating) conditions is done
to show that the nature of convective onset is not different from that for
insulating walls. As isothermal conditions are maintained for the basic state,
 the temperature perturbation vanishes at the top and bottom. As the
horizontal ($x$) direction mimics the azimuthal ($\phi$) direction
in cylindrical polar coordinates, periodic conditions
are set at the side walls. The boundary conditions on $z$ are implemented
as follows:
\begin{align}
u_z=D_z^2 u_z = D_z \omega_z =0 \ \,\, &\text{at} \ z=0, 1 \ \text{(stress-free)};\\
u_z=D_z u_z =\omega_z =0 \ \,\, &\text{at} \ z=0, 1 \ \text{(no-slip)};\\
j_z=0 \,\,&\text{at} \ z=0, 1 \ \text{(both walls insulating)};\\
b_z = D_z j_z=0 \,\,&\text{at} \ z=0 \ \text{(bottom wall conducting)};\\
\theta =0 \,\, &\text{at} \ z=0, 1. 
\end{align}

\subsection{Method of solution and benchmarks}
\label{bench}
The stationary onset of magnetconvection with a laterally varying field 
is studied for the parameters 
$E=5 \times 10^{-4}-5 \times 10^{-7}$, $\varLambda=0-1$ and
$q=1$. The generalized eigenvalue problem $\bm {AX}=\lambda \bm {BX}$, 
where $\lambda=Ra$ is solved
using Matlab. For the set of equations 
(\ref{mom2})--(\ref{current2}), the matrices and
their elements are presented in Appendix \ref{app1}. 
A spectral collocation method
that uses Chebyshev differentiation in $z$ and 
Fourier differentiation in $x$ is used
to resolve the eigenfunctions in two dimensions. For problems
with variable coefficient terms (as in this study),
 the spectral collocation method
uses simple matrix multiplication in physical space to treat the terms,
whereas a pure spectral method would have resulted in convolution sums for
such terms that are algebraically complex \citep{peyret}. The drawback
of the collocation method, however, is that the differentiation matrices
are dense, making computations memory-intensive \citep{muite,12benhadid}.
The construction of the Fourier and Chebyshev
differentiation matrices
follows a standard approach, and is given in Appendix \ref{app1}
for completeness. For
$E=5 \times 10^{-7}$ with stress-free boundaries, grid independence is secured
with $18$ points in $z$ and $210$ points in $x$, 
so the non-zero elements of ${\bm A}$
and ${\bm B}$ are of size $(18 \times 210)^2$. 

\begin{figure}
\begin{center}
\resizebox{14 cm}{!} {\includegraphics{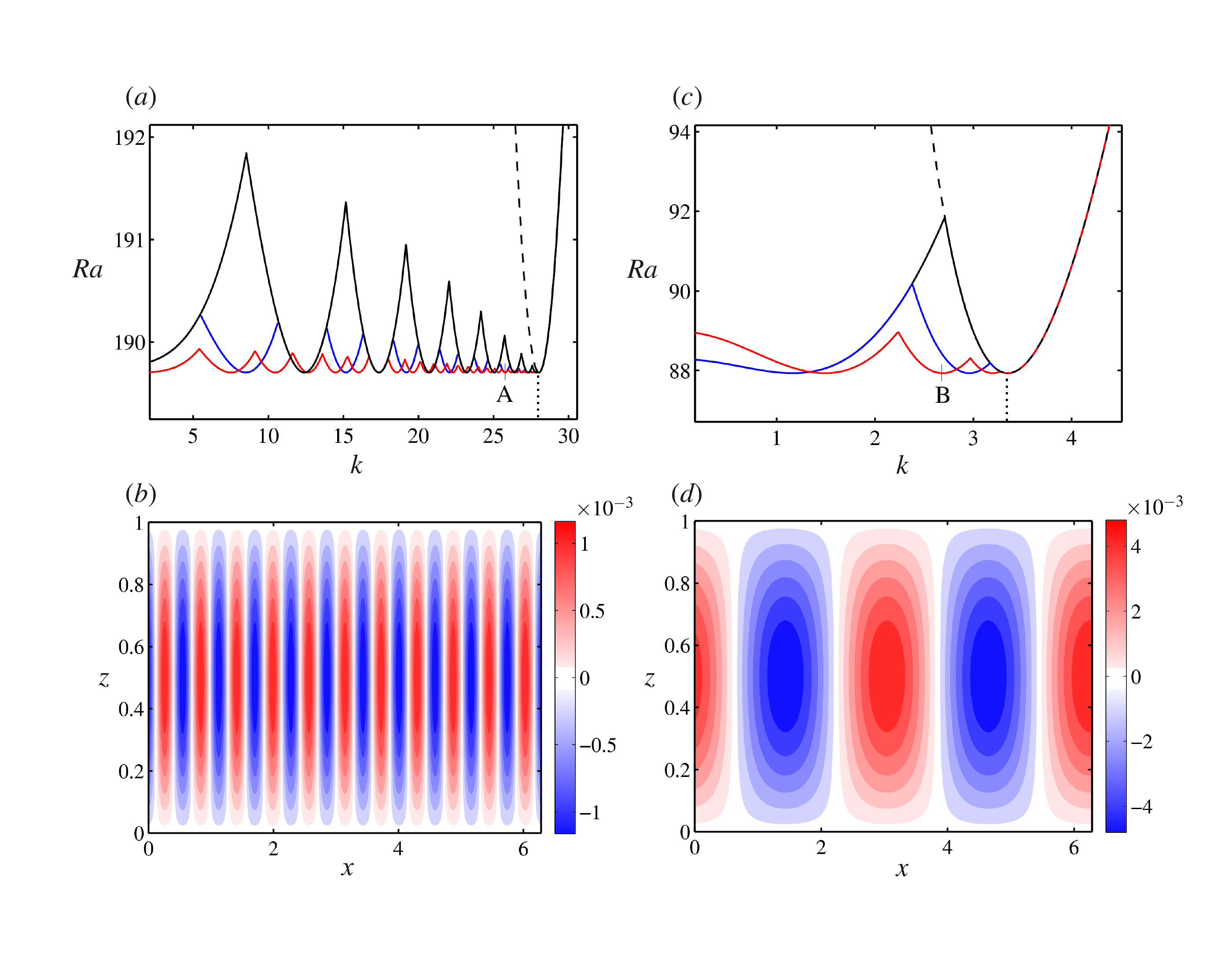}}\\
\caption{(\emph{a}) Neutral stability curves for non-magnetic convection
in a  plane layer of finite aspect ratio for 
$E= 1 \times 10^{-4}$ and stress-free
$z$-boundaries. The cases shown are
$L_x=2$ (black), $L_x=4$ (blue) and $L_x= 2\pi$ (red). 
(\emph{b}) Axial velocity ($u_z$) for the unstable mode marked \lq A' 
($L_x=2\pi, k=25.78$) in (\emph{a}). (\emph{c}) Neutral stability curves for
magnetoconvection at $E= 1 \times 10^{-4}$, $\varLambda=0.5$ and $q=1$, for the 
same cases (and line styles) as (\emph{a}). (\emph{d}) $u_z$ for the 
unstable mode marked \lq B' 
($L_x=2\pi, k=2.69$) in (\emph{c}). The dashed lines in
 (\emph{a}) and  (\emph{c}) are the neutral curves for the infinite plane layer.
The layer depth $L_z=1$ in all cases. }
\label{fig5}
\end{center}
\end{figure}

Figure \ref{fig5} shows the existence of a finite number of
equally unstable $y$-wavenumbers ($k$) at the onset of convection 
in a plane layer of finite aspect ratio.
 The exact number and values of the unstable wavenumbers
are predictable for a given horizontal lengthscale $L_x$ 
(Appendix \ref{app2}). 
As $L_x \to \infty$, the number of unstable wavenumbers 
would become infinite. 
 For a given $L_x$,
\begin{equation}
a^2 = \biggl(\frac{2m \pi}{L_x}\biggr)^2 + k^2, \nonumber
\end{equation}
where $m$ is the $x$-wavenumber and $a$ is the
resultant wavenumber.
Consequently, the last unstable $y$-wavenumber 
coincides with the critical wavenumber 
$a_c$ for the classical one-dimensional plane layer 
of infinite horizontal extent \citep{61chandra}.
For $E= 1 \times 10^{-4}$, non-magnetic convection gives
$a_c=28.02$; 
and for $L_x=2\pi$, the axial velocity $u_z$
at the unstable wavenumber marked A in figure \ref{fig5}(\emph{a})
shows 11 pairs of rolls (figure \ref{fig5}\emph{b}), 
consistent with the fact that 
$m_c=\sqrt{28.02^2-25.78^2} \approx 11$. In a similar way, convection 
under a uniform axial magnetic field for $E=1 \times 10^{-4}$
and $\varLambda=0.5$ at the point B in
figure \ref{fig5}(\emph{c}) produces 2 pairs of
rolls (figure \ref{fig5}\emph{d}) because $a_c=3.35$ and 
$m_c=\sqrt{3.35^2-2.69^2} \approx 2$.

\subsection{Onset of convection under a laterally varying magnetic field}      
\label{onset}
\begin{figure}
\begin{center}
\resizebox{10 cm}{!} 
{\includegraphics{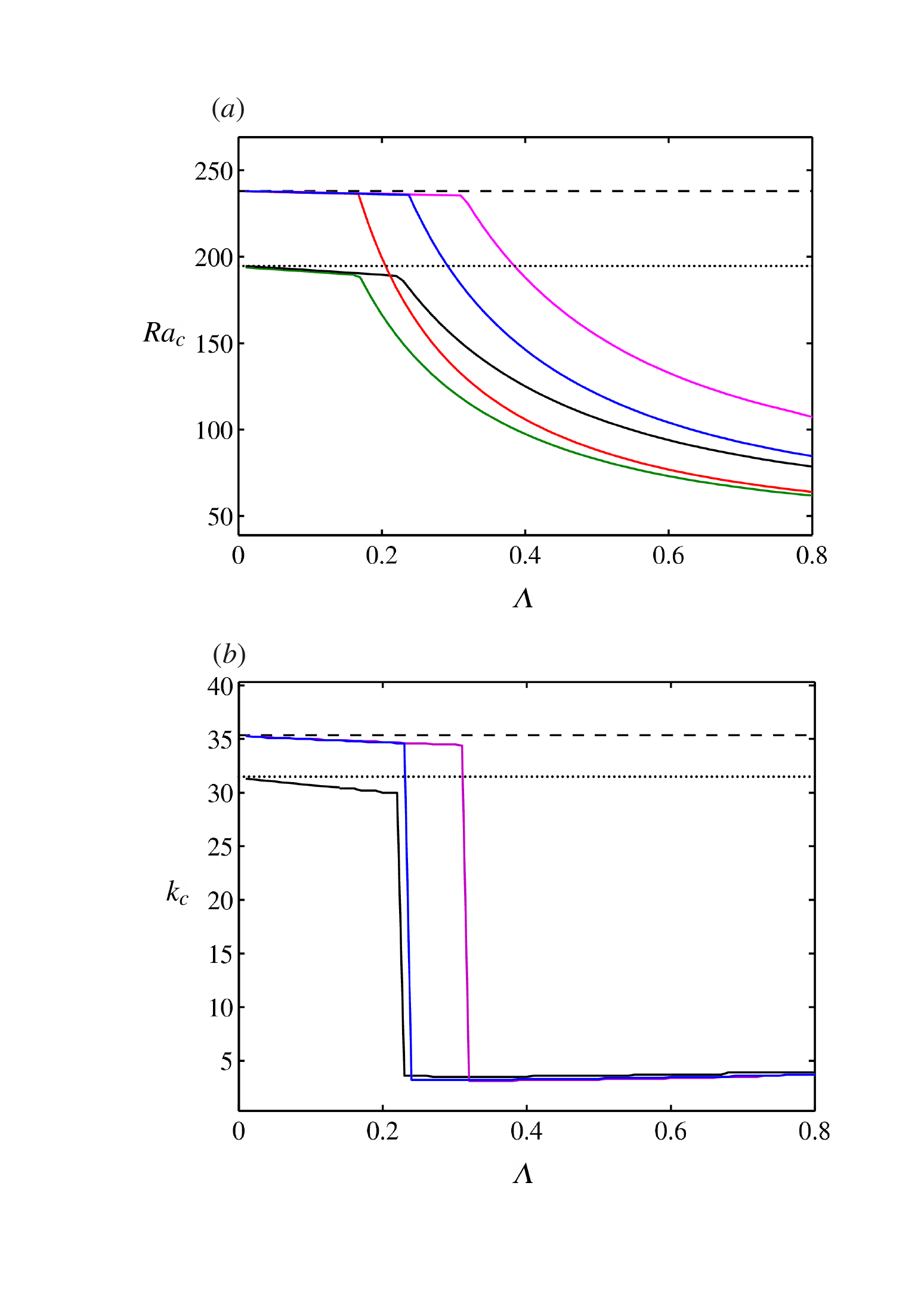}}\\
\caption{$Ra_c$\,--\,$\varLambda$ and $k_c$\,--\,$\varLambda$ 
regime diagrams for $E=5 \times 10^{-5}$.
The reference values for the non-magnetic case ($\varLambda=0$) 
are given by the 
horizontal dashed (stress-free) and
dotted (no-slip) lines. The uniform magnetic
field cases are given by red (stress-free and insulating) and green 
(no-slip and insulating) lines. The inhomogeneous magnetic field 
cases are given by blue (stress-free and insulating),
 magenta (stress-free and mixed) and black (no-slip and insulating) lines.
}
\label{fig6}
\end{center}
\end{figure}

\begin{figure}
\begin{center}
\resizebox{14 cm}{!} {\includegraphics{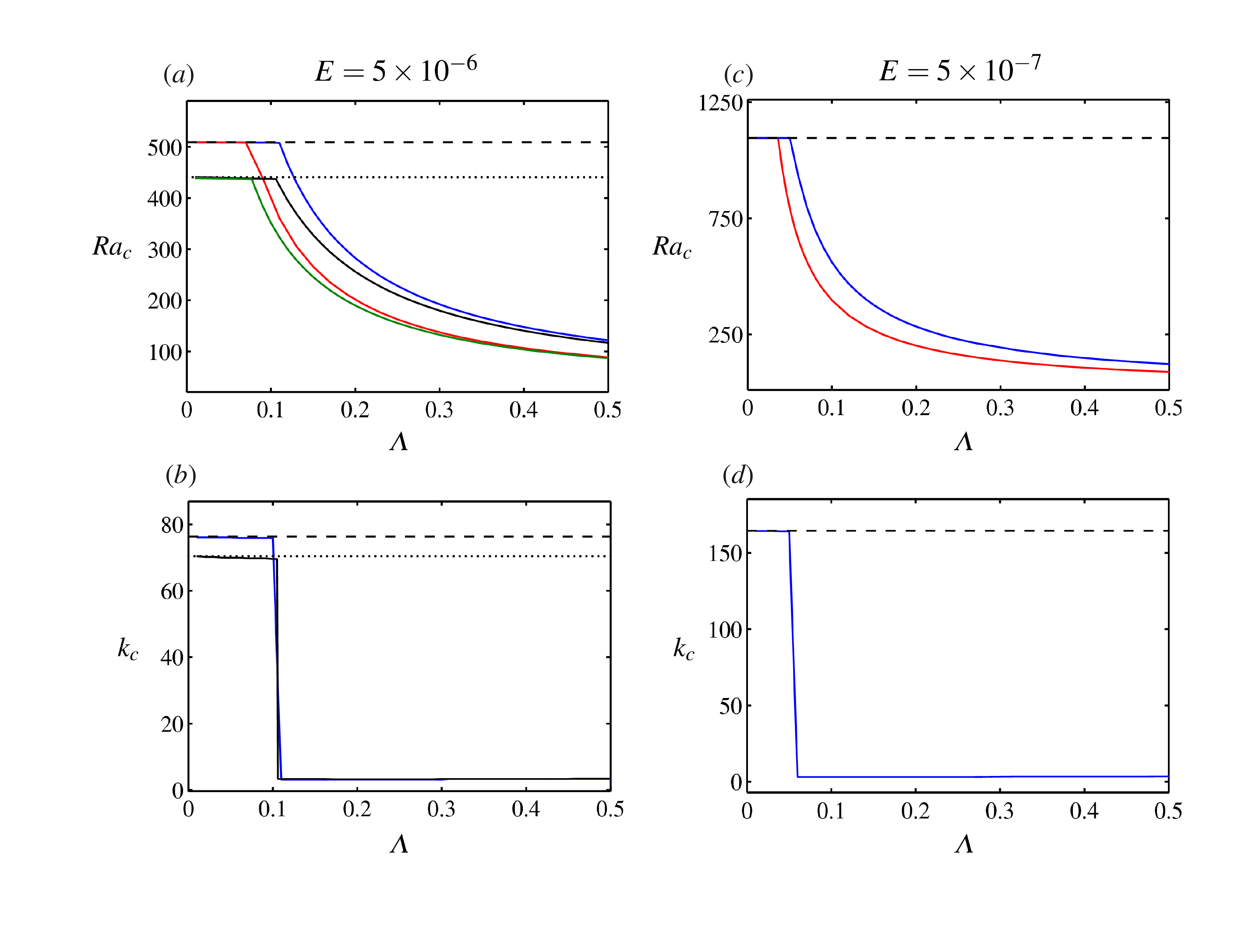}}\\
\caption{$Ra_c$\,--\,$\varLambda$ and $k_c$\,--\,$\varLambda$ 
regime diagrams for $E=5 \times 10^{-6}$ and $E=5 \times 10^{-7}$.
The reference values for the non-magnetic case ($\varLambda=0$) 
are given by the 
horizontal dashed (stress-free) and
dotted (no-slip) lines. The uniform magnetic
field cases are given by red (stress-free and insulating) and green 
(no-slip and insulating) lines. The inhomogeneous field 
cases are given by blue (stress-free and insulating)
and black (no-slip and insulating) lines.
}
\label{fig7}
\end{center}
\end{figure}

We investigate marginal-state convection in a plane 
layer of depth $L_z=1$ and
horizontal lengthscale $L_x=4$ permeated by an inhomogeneous axial ($z$)
magnetic field of the form (\ref{bprofile}) giving strong localization
of the field (see figure \ref{fig4}\emph{b}). The background value of the field
is small compared to its peak value but not zero, in line with the 
axial field distribution within the tangent cylinder in rapidly
rotating dynamo simulations. Figures \ref{fig6} and \ref{fig7}
 give the critical Rayleigh
number ($Ra_c$) and wavenumber ($k_c$) diagrams for this field distribution, 
with the reference
states for non-magnetic convection and homogeneous magnetic field provided for
comparison. The critical wavenumber for the homogeneous field is not
shown because its value is not unique, as noted earlier in \S \ref{bench}
(although the critical resultant wavenumber $a_c$ is unique).
For the laterally varying field, 
$Ra_c$ follows the same trend as for the homogeneous field, being approximately 
constant in the large-wavenumber viscous branch and then falling steeply 
in the small-wavenumber magnetic branch. The field inhomogeneity, 
however, displaces the
viscous--magnetic mode transition point to a higher Elsasser number. 
Changing the mechanical
and electromagnetic $z$-boundary conditions does not alter the basic properties
of the regime diagrams, although the numerical values of 
$Ra_c$ and $k_c$ differ
from one condition to the other. While the viscous branches 
for insulating and mixed 
(top insulating and bottom perfectly conducting) electromagnetic conditions 
largely overlap,
the use of mixed  conditions moves the viscous--magnetic transition
further to the right (compare the blue and magenta lines in 
figure \ref{fig6}\emph{a} and \emph{b}).
Table \ref{table1} (for stress-free conditions) and table \ref{table2}
(for no-slip conditions) present selected values of the critical parameters
spanning the two branches of instability. 
A notable property of onset in the magnetic
mode is that $Ra_c$ and $k_c$ are nearly independent of the Ekman number $E$,
in agreement with the classical picture of onset in an infinite
plane layer under a uniform magnetic field \citep[see, for example,
figure 3 in][]{15kelig}. In this regime, the critical temperature
gradient for convection is independent of viscosity, and it is
the magnetic field via the Lorentz force that breaks the Taylor-Proudman
constraint to set up convection in the fluid layer.
 
\begin{table}	
	\begin{center}
	\begin{tabular}{ c c c c c c c c c c c}
                \multicolumn{4}{c@{}}{$E=5\times10^{-5}$} &
		\multicolumn{3}{c@{}}{$E=5\times10^{-6}$} &
		\multicolumn{4}{c@{}}{$E=5\times10^{-7}$} \\
		\multicolumn{4}{c}{(Stress-free)} &
		\multicolumn{3}{c}{(Stress-free)} &
		\multicolumn{4}{c}{(Stress-free)} \\

		$\varLambda$ & $Ra_{c}$ & $k_{c}$ &&  $\varLambda$ & $Ra_{c}$ & $k_{c}$ && 
 $\varLambda$ & $Ra_{c}$ & $k_{c}$\\ [0.5ex]
		\hline
		0     & 237.91 & 35.35  & \hspace{0.2cm}   &  0   & 509.41     & 76.27 
& \hspace{0.2cm}  &    0     & 1096 & 164.38   \\
		0.001 & 237.83 &  35.3  &&  0.001 & 509.37 & 76.25 &&  0.001 & 1095.97 & 164.3   \\
		0.01  & 237.82 & 35.3  && 0.01 & 509.30  & 76.1  &&  0.01 & 1095.90 &  164.3   \\
		0.04  & 237.57 & 35.1 && 0.04   & 509.01     & 76.1 && 0.04  & 1095.72 & 164.1  \\
		0.1  & 237.03 & 35    && 0.1   & 508.43     & 75.9  &&    0.05  & 1095.61 & 164.1 \\
		0.2  & 236.10 & 34.7  && 0.11*  & 507.73     & 3.1; 75.9 &&  0.051  & 1095.22 & 3.06 \\
		0.22  & 235.90 & 34.6  && 0.12    & 465.91    & 3.08 &&   0.1  & 560.33 & 3.07  \\
		0.238*  & 235.70 & 3.2; 34.6 &&  0.15 & 374.14  & 3.09 &&   0.15  & 375.23 & 3.08 \\
		0.239  & 233.44 & 3.16   && 0.2    & 282.61     & 3.11 &&    0.2  & 283.37 & 3.1  \\
		0.3  & 189.01 & 3.2   &&  0.3    & 191.92     & 3.17 &&     0.3  & 192.20 & 3.16\\
		0.5  & 120.58 & 3.35       & &  0.5    & 121.62     & 3.35 &&    0.5  & 121.73 & 3.35\\
	        0.6  & 104.29 & 3.5     &&  0.6   & 105.01     & 3.48 & &   0.6  & 105.08 & 3.48\\
		0.8  & 85.28 & 3.8        & &  0.8   & 85.68     & 3.78 & &     0.8  & 85.72 & 3.78\\
		1.0  & 75.33 & 4.15      & &  1.0   & 75.57     & 4.13 & &     1.0  & 75.60 & 4.13\\
	\end{tabular}
        \end{center}
	\caption{Rayleigh numbers ($Ra_c$) and wavenumbers ($k_c$) for marginal
		state (critical) convection, computed for Elsasser numbers ($\varLambda$) 
		in the range 0--1 and stress-free $z$-boundaries. 
*Denotes the viscous--magnetic
mode transition point.}
	\label{table1}
\end{table}

\begin{table}
	\centering
	\begin{tabular}{ c c c c c c c }
		\multicolumn{4}{c}{$E=5\times10^{-5}$} &
		\multicolumn{3}{c}{$E=5\times10^{-6}$} \\
		\multicolumn{4}{c}{(No-slip)} &
		\multicolumn{3}{c}{(No-slip)} \\
		
		$\varLambda$ & $Ra_{c}$ & $k_{c}$ & \hspace{0.2cm} & 
$\varLambda$ & $Ra_{c}$ & $k_{c}$\\ [0.5ex]
		\hline
		0    & 194.65   & 31.3 & &     0     & 440.82 & 70.5   \\
		0.001  & 194.61 &  31.3 & &     	0.001 & 440.81 & 70.4   \\
		0.01   & 194.56 & 31.3  & &    0.01 & 440.80 &  70.4  \\
		0.04   & 193.80 & 31.1  & &     0.04  & 439.73 & 70  \\
		0.1   & 192.21  & 30.7  & &   0.1  & 437.45 & 69.6 \\
		0.15   & 190.87 & 30.4 & &   0.107*  & 436.91 & 3.3; 69.6 \\
		0.2   & 189.50  & 30   & &    0.11  & 424.03 & 3.3  \\
		0.225*   & 188.78  & 3.6; 30 & & 0.2  & 256 & 3.2 \\
		0.23   & 185.95  & 3.6 & & 0.3  & 179.78 & 3.3  \\
		0.3   & 154.15  & 3.5 & &  0.4  & 140.42 & 3.3  \\
		0.5   & 106.54  & 3.5 & &     0.5  & 116.83 & 3.4\\
		0.6   & 93.99  & 3.6 & &    0.6  & 101.41 & 3.5 \\
	\end{tabular}
	\caption{Rayleigh numbers ($Ra_c$) and wavenumbers ($k_c$) 
for marginal state (critical) convection, computed for 
Elsasser numbers ($\varLambda$) in the range 0--0.6 and no-slip $z$-boundaries.
 *Denotes the viscous--magnetic
mode transition point.}
	\label{table2}
\end{table}

\begin{figure}
\begin{center}
\resizebox{14 cm}{!} {\includegraphics{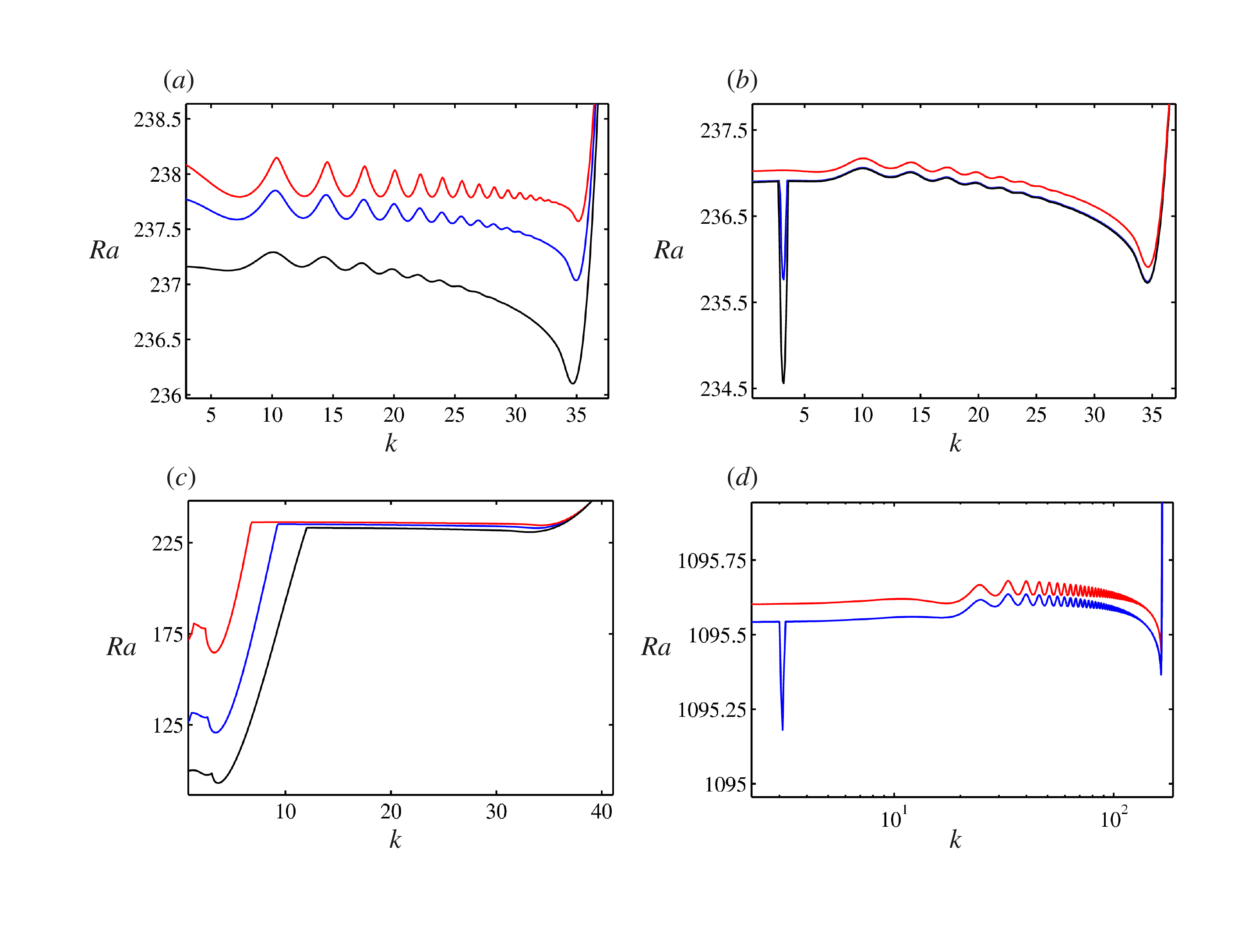}}\\
\caption{Neutral stability curves for different magnitudes of the 
imposed inhomogeneous magnetic field 
in figure \ref{fig4}(\emph{b}) and stress-free, 
electrically insulating $z$-boundaries. Two Ekman numbers
are analyzed, $E=5 \times 10^{-5}$ (\textit{a--c}) and $E=5 \times 10^{-7}$ (\textit{d}).
(\textit{a}) $\varLambda=0.04$ (red), $\varLambda=0.1$ (blue), $\varLambda=0.2$ (black).
(\textit{b}) $\varLambda=0.22$ (red), $\varLambda=0.238$ (blue), 
$\varLambda=0.239$ (black).
(\textit{c}) $\varLambda=0.3$ (red), $\varLambda=0.5$ (blue), $\varLambda=0.7$ (black).
(\textit{d}) $\varLambda=0.04$ (red), $\varLambda=0.051$ (blue).}
\label{fig8}
\end{center}
\end{figure}
Figure \ref{fig8} presents the 
neutral stability curves extracted from various points
on the regime diagrams for two Ekman numbers. 
For $E=5 \times 10^{-5}$, the laterally varying magnetic 
field of small strength $\varLambda=0.04$ forces a unique mode of instability 
($k_c=35.1$), even as the vestige of the multiple-wavenumber, 
non-magnetic solution is visible in the oscillations
of the curve (figure \ref{fig8}(\emph{a}), red line). 
The amplitude of the oscillations
decreases with increasing Elsasser number, and for $\varLambda=0.238$ 
(blue line in figure \ref{fig8}\emph{b}),
the magnetic mode of onset ($k_c=3.2$) appears at the same Rayleigh number 
as the viscous mode
($k_c=34.6$). For $\varLambda=0.239$, the magnetic mode overtakes 
the viscous mode as the most unstable.
As $\varLambda$ is increased further, $Ra_c$ progressively 
decreases but $k_c$ remains approximately constant
(figure \ref{fig8}\emph{c}). The viscous--magnetic mode transition at 
$E=5 \times 10^{-7}$ takes place over a very
narrow range of Elsasser numbers, with $\varLambda=0.05$ showing 
viscous onset ($k_c=164.1$) and
$\varLambda=0.051$ showing magnetic onset ($k_c=3.06$; 
blue line in figure \ref{fig8}\emph{d}). 
(The logarithmic $x$-axis scale of figure \ref{fig8}(\emph{d}) 
shows the 
large scale separation between the viscous and magnetic 
modes clearly). The large-wavenumber
oscillations still exist, although with the laterally varying magnetic field
these are never the most unstable modes.

\begin{figure}
\begin{center}
\resizebox{13.5 cm}{!} 
{\includegraphics{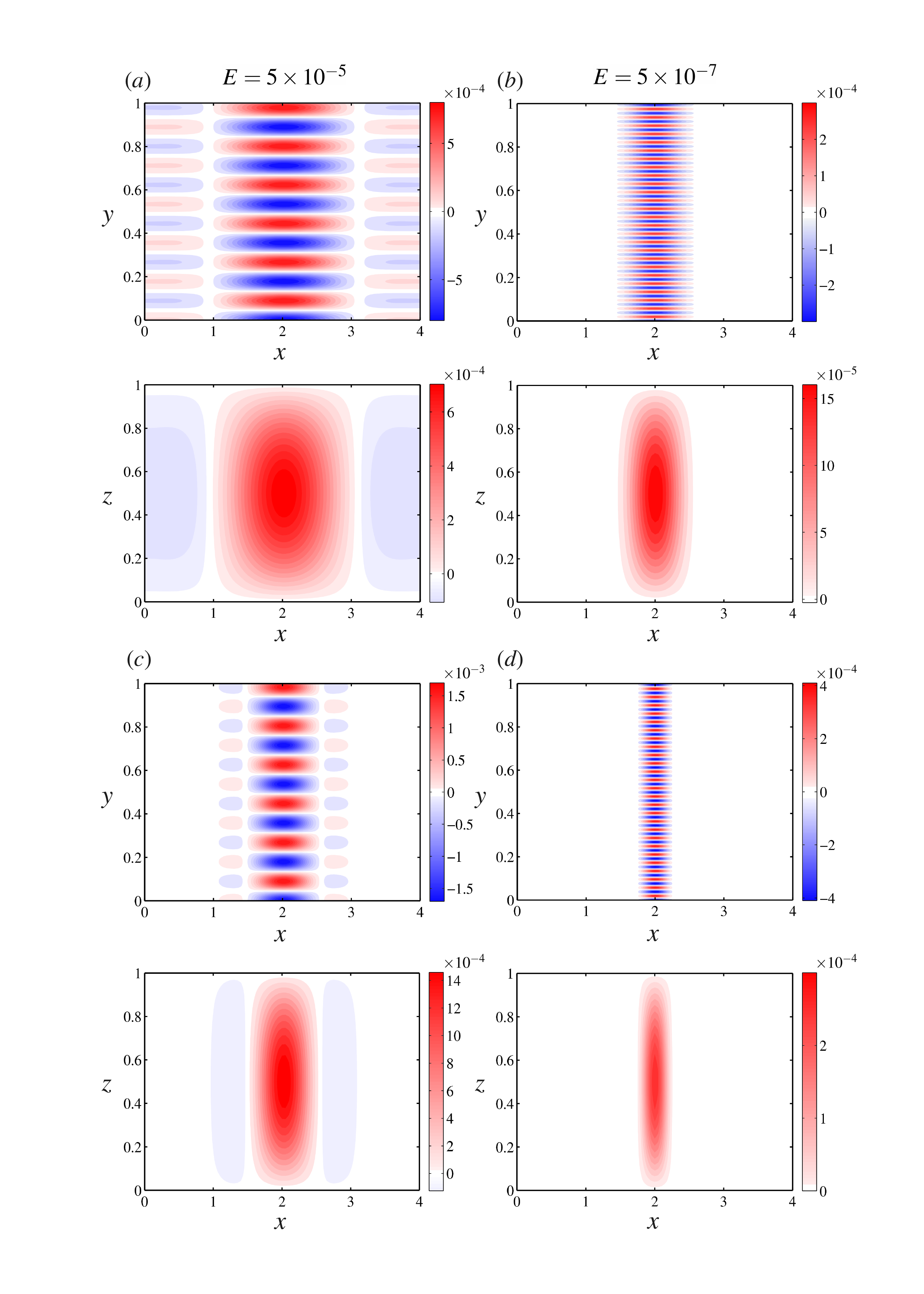}}\\
\caption{Contour plots of the axial velocity 
$u_z$ for two Ekman numbers ($E$) at onset of magnetoconvection
on the $(x,z)$ and $(x,y)$ planes, with a 
restricted $y$-range of $[0,1]$. (\emph{a}) and
(\emph{b}): $\varLambda=10^{-3}$.  
(\emph{c}) and (\emph{d}): $\varLambda=0.04$. The
$z$-boundaries are stress-free and electrically insulating.}
\label{fig9}
\end{center}
\end{figure}

\begin{figure}
\begin{center}
\resizebox{13.5 cm}{!} 
{\includegraphics{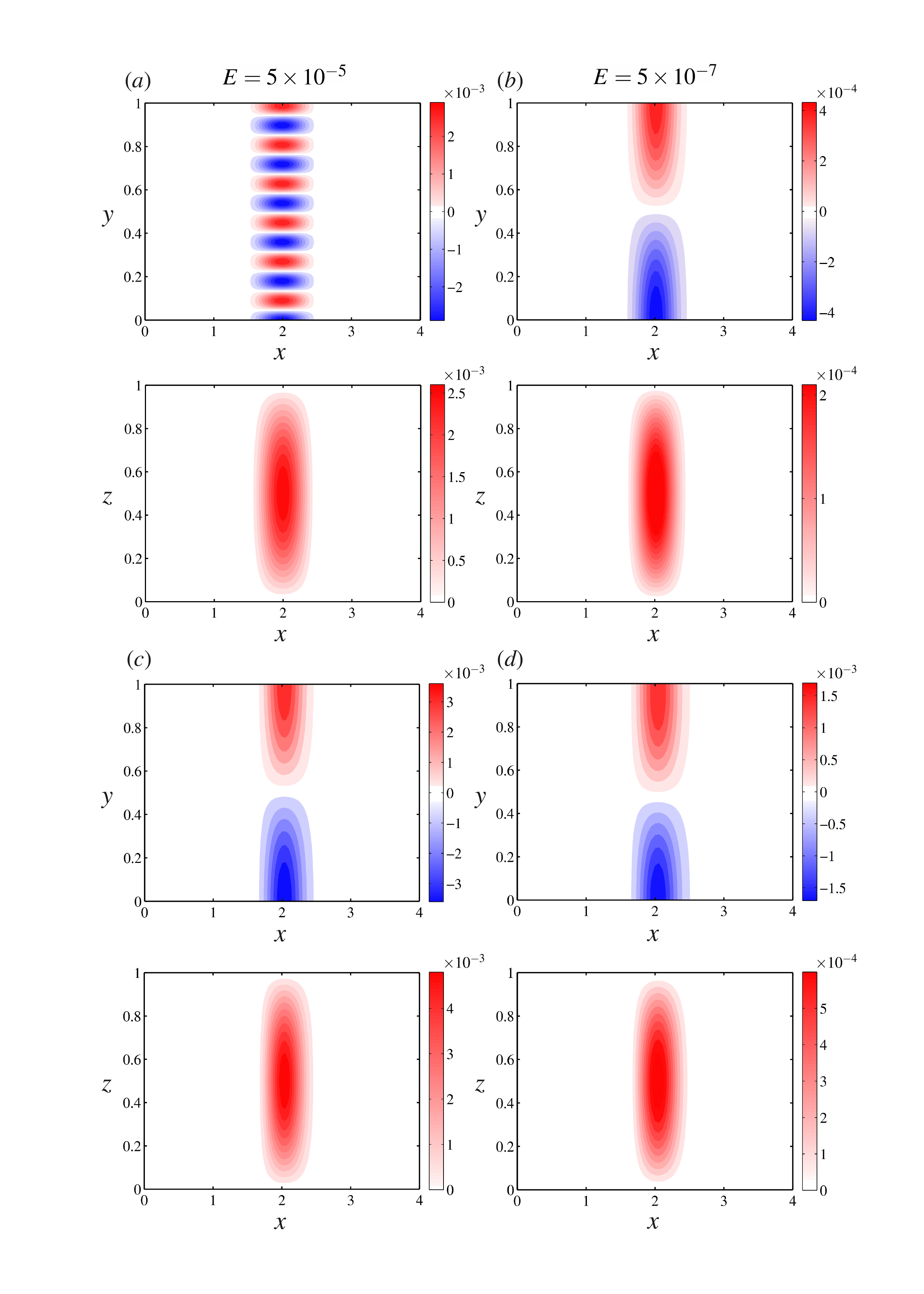}}\\
\caption{Axial velocity $u_z$ for two Ekman numbers 
($E$) at onset of magnetoconvection
on the $(x,z)$ and $(x,y)$ planes.  (\emph{a}) and
(\emph{b}): $\varLambda=0.1$.  (\emph{c}) and (\emph{d}): $\varLambda=0.3$.
The $z$-boundaries are stress-free and electrically insulating.}
\label{fig10}
\end{center}
\end{figure}

Figures \ref{fig9} and \ref{fig10} show the axial velocity 
$u_z$ at convective onset for
two Ekman numbers. (Both $u_z$ and $\theta$ have identical
structures.) The main idea that comes out of this study 
is that convection takes the
form of isolated plumes under a laterally varying
magnetic field even when the smallest lengthscale of the flow 
is controlled by viscosity.
For a small field strength $\varLambda=10^{-3}$, a unique mode of
instability develops where convection is concentrated in the 
neighbourhood of the peak magnetic field at $x=2$ 
(figures \ref{fig9}\textit{a},\textit{b}). 
(It has been confirmed that moving the peak location of the imposed 
field by changing
the constant $c$ in equation (\ref{bprofile}) also
 moves the location of convection).
The large number of convection cells stacked 
in the $(x,y)$ plane points to the viscous mode
of onset. Convection here is magnetically confined, yet its
smallest lengthscale is viscously controlled. 
The magnetic field can therefore help overcome the
Taylor-Proudman constraint and set up localized 
convection while not significantly
changing the wavenumber of convection from its non-magnetic value. 
It is notable that the field-induced localization 
is more pronounced at the 
lower Ekman number:
for $\varLambda=10^{-3}$ and $\varLambda=0.04$, the rolls at 
$E=5 \times 10^{-7}$ are appreciably thinner than
at $E=5 \times 10^{-5}$, 
although the imposed magnetic field profile 
is the same in both cases
(figures \ref{fig9}\textit{a} and \textit{b};
 \ref{fig9}\textit{c} and \textit{d}). 
The formation of thin, yet 
isolated plumes has direct relevance to convection within
the tangent cylinder in rapidly rotating  
spherical dynamo simulations (\S \ref{dynamo})
where similar structures are noted. As the field strength 
is increased further to
$\varLambda=0.1$, convection at $E=5 \times 10^{-5}$ is 
still in the large-wavenumber
viscous branch, whereas convection at $E=5 \times 10^{-7}$ has crossed over 
to the small-wavenumber magnetic branch (figures \ref{fig10}(\textit{a}) 
and (\textit{b}) and table \ref{table1}).
From figures \ref{fig10}(\textit{c}) and (\textit{d}) ($\varLambda=0.3$), 
it is noted that
the small-wavenumber convection at both Ekman numbers is almost identical
in structure, consistent with the fact that the critical parameters
($Ra_c$, $k_c$) are independent of
Ekman number beyond the viscous--magnetic transition point 
\citep[e.g.][]{15kelig}. 
\begin{figure}
\begin{center}
\resizebox{14 cm}{!} {\includegraphics
{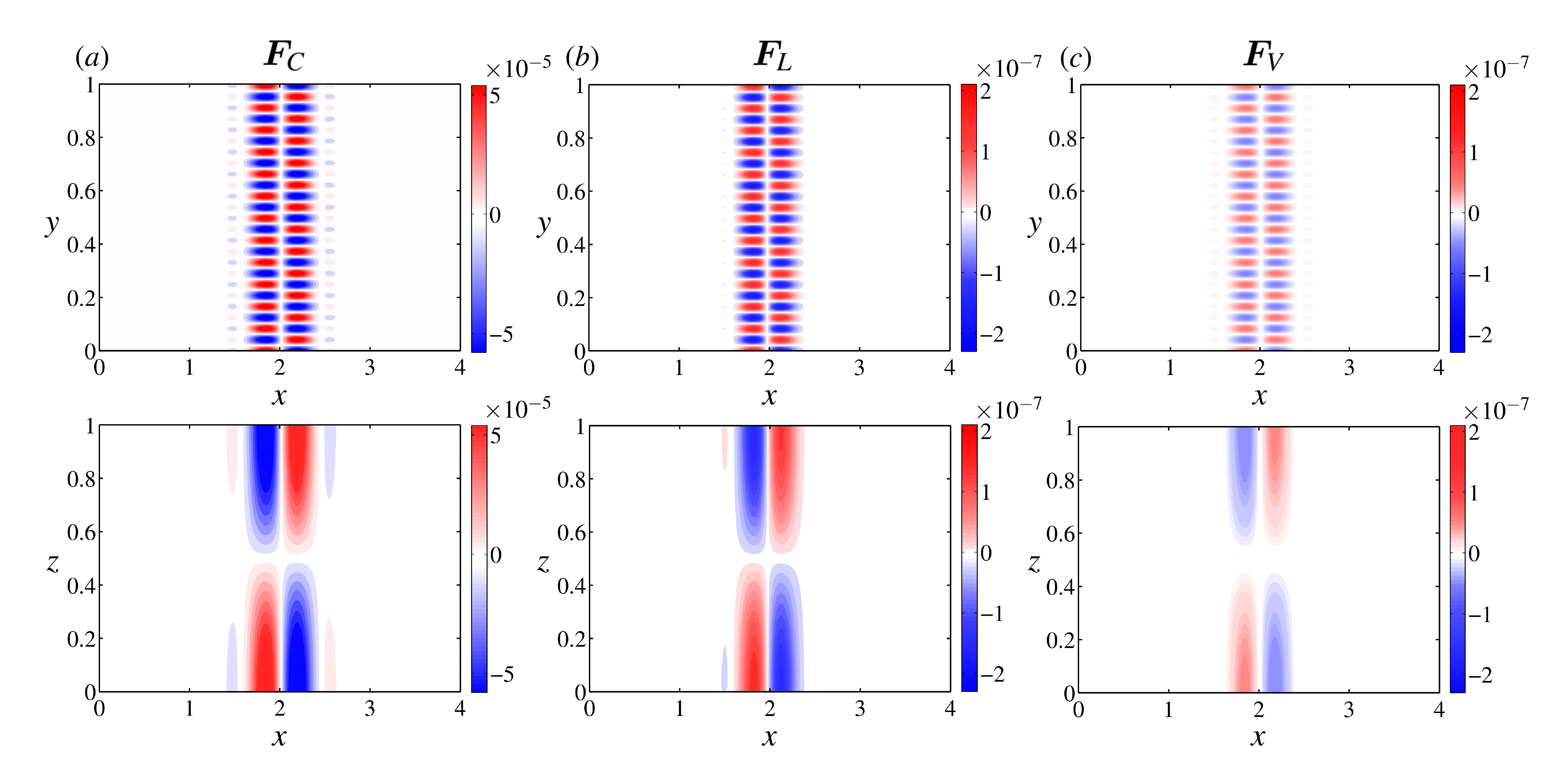}}\\
\caption{Contour plots of the $x$-components of the Coriolis force $\bm{F}_C$,
Lorentz force $\bm{F}_L$ and viscous force $\bm{F}_V$ on the $(x,z)$
and $(x,y)$ planes. The $(x,z)$ plane is shown at $y=0$.
The parameters are $E= 5 \times 10^{-6}$ and 
$\varLambda=0.08$. The $z$-boundaries are stress-free and electrically
insulating.
}
\label{fig11}
\end{center}
\end{figure}

Figure \ref{fig11} shows the $x$-components of the Coriolis, Lorentz and
viscous forces (denoted by subscripts $C$, $L$ and $V$ respectively) in
the momentum equation (\ref{mom1}). (In this model, the pressure
gradient is not solved for). Here $E=5 \times 10^{-6}$,
for which $\varLambda=0.08$ gives onset in the viscous
branch (table \ref{table1}). 
The $x$-component of $\bm{F}_C$ gives $u_y$. From the
plots of the Lorentz and viscous forces shown on the same
colour scale (figure \ref{fig11}\emph{b,c}), it is
inferred that the Lorentz force, whose magnitude is $\approx 5$ times
that of the viscous force, is influential in overcoming
the Taylor-Proudman constraint and setting up convection.
Interestingly, $\varLambda \bigl[(\bm{\nabla} \times \bm{b})
  \times \bm{B}_0 \bigr]$ makes the dominant contribution
to the Lorentz force
while $\varLambda \bigl[(\bm{\nabla} \times
  \bm{B}_0) \times \bm{b} \bigr]$ is slightly
smaller in magnitude than the viscous force $E \nabla^2 \bm{u}$
(see equation \ref{mom1}).
At onset in the magnetic branch ($\varLambda=0.3$),
the peak value of $\bm{F}_L$ is
three orders of magnitude larger than that of $\bm{F}_V$ and
only one order of magnitude smaller than that of $\bm{F}_C$, which
emphasizes the well-known role of the magnetic field in
overcoming the rotational constraint.
\begin{figure}
\begin{center}
\resizebox{14 cm}{!} {\includegraphics
{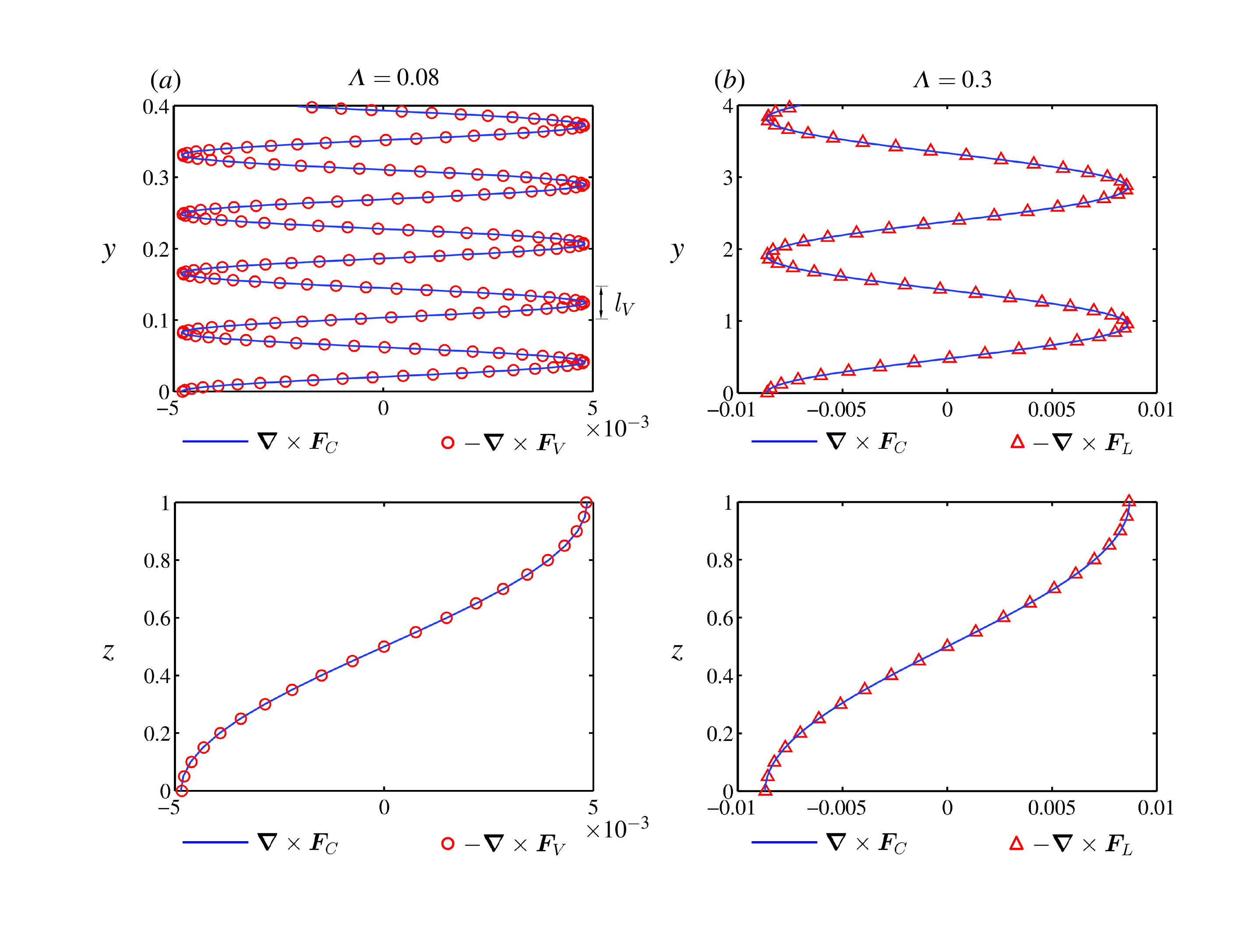}}\\
\caption{Line plots at $x=2$ showing the principal balance of terms in the
  $z$-vorticity equation for two values of $\varLambda$ that
  represent the viscous and magnetic modes of onset. The $z$-variation
is shown at $y=0$.
  The Ekman number $E=5 \times 10^{-6}$. The $z$-boundaries
  are stress-free and electrically insulating.
}
\label{fig12}
\end{center}
\end{figure}

The mode of convective onset is reflected in the principal balance of terms
in the $z$-vorticity equation (\ref{vort2}). For
$\varLambda=0.08$, $\bm{\nabla}\times {\bm F}_C$ closely
matches $- \bm{\nabla}\times {\bm F}_V$ (figure \ref{fig12}\emph{a}),
whereas for $\varLambda=0.3$, $\bm{\nabla}\times {\bm F}_C$ closely
matches $- \bm{\nabla}\times {\bm F}_L$ (figure \ref{fig12}\emph{b}).
The Lorentz force term has a negligible contribution to 
the balance in the former case while the
viscous force term has a negligible effect in the latter. 
The $y$-axis range for the two cases is chosen such
that the difference in the lengthscale of
convection is clear. Although
the large-scale magnetic mode of onset
in figure \ref{fig12}(\emph{b})
is well understood, the role of the Lorentz force
in setting up convection
at the small viscous lengthscale $l_V$ (figure \ref{fig12}\emph{a},
upper panel) has not received much attention
in the literature.

Increasing the background magnetic field intensity
relative to the peak value reduces the lateral inhomogeneity
of the field. For moderate lateral variation, obtained by
progressively increasing $a_0$ in the mean 
field profile (\ref{bprofile}), convection at onset
retains the structure of isolated rolls centered at
the location of the peak field. A weak
lateral variation ($a_0 \sim 0.8$), however,
gives rise to a cluster of rolls in the viscous mode
whose intensity decays from the centre ($x=2$)
towards the periphery. As the
lateral variation goes to zero, 
the solution would tend to that for a homogeneous
magnetic field (\S \ref{bench}). 

In summary, a 
laterally varying magnetic field acting on a rotating fluid layer
gives rise to a unique mode of instability 
where convection follows
the path of the peak field. This localized excitation
of convection is consistent with the
idea that the magnetic field generates helical
fluid motion in regions that are otherwise
quiescent \citep{11sreeni}, although here it is shown
that the flow lengthscale at onset could be 
viscously or magnetically controlled. The critical 
Rayleigh number for magnetic convection increases with 
decreasing Ekman
number in the viscous branch of onset, 
whereas it is nearly independent of
Ekman number in the magnetic branch. The width
of the convection zone decreases with decreasing 
Ekman number in the viscous
branch, whereas it is nearly independent of 
Ekman number in the magnetic branch.

\subsection{Implications for onset of localized convection 
within the tangent cylinder}
\label{implications}
From a comparison between the plane layer linear magnetoconvection model
and the spherical shell dynamo simulations, the following points
are noted:
\begin{enumerate}
\item The Rayleigh number for onset of localized
convection (in the form of an isolated plume)
within the tangent cylinder ($Ra=190$ for $E=5 \times 10^{-5}$
and $Ra=438$ for $E=5\times 10^{-6}$, with no-slip boundaries) 
lies on the viscous branch
of onset in the plane layer magnetoconvection model ($Ra_c=194.6$--$188.8$
 for $E=5 \times 10^{-5}$ and $Ra_c=440.8$--$436.9$
 for $E=5\times 10^{-6}$, with no-slip boundaries). 
This agreement between the plane layer
and the spherical dynamo Rayleigh numbers rests on two
factors: (a) The dominant magnetic field within the 
tangent cylinder is axial; and (b) the formation of 
localized convection within the tangent cylinder
is practically unaffected by the curvature of the bounding
walls. It is notable that the plane layer model does not
predict the critical Rayleigh number for non-magnetic
convection in the tangent cylinder, as the sloping walls
allow uniformly distributed convection at a lower
Rayleigh number (e.g. $Ra \sim 140$ for $E=5 \times 10^{-5}$).

\item The width of an isolated plume within the tangent cylinder 
markedly decreases with 
decreasing Ekman number (figure \ref{fig3}\emph{b,e}), an effect
that is noted only in the viscous branch of onset in the
magnetoconvection model (figure \ref{fig9}\emph{c,d}). Had
the plume formed in the magnetic branch of onset, its width should 
have been nearly independent of the Ekman number
 (figure \ref{fig10}\emph{c,d}).
\end{enumerate}  

The close agreement between the 
dynamo and viscous-branch magnetoconvection
Rayleigh numbers notwithstanding, the critical 
Elsasser number $\varLambda_z$ 
in the simulation is much higher than that
in the linear magnetoconvection model. 
For example, in the dynamo simulation at $E=5 \times 10^{-6}$,
plume onset occurs for $\varLambda_z \approx 16.8$
while $Ra$ does not depart much
from its nonmagnetic value with no-slip
boundaries, $440.82$ (see tables \ref{table0} and \ref{table2}).  
This striking difference in $\varLambda_z$ 
between the spherical shell tangent cylinder
and the plane layer model
is because of the naturally occurring
axial variation of $B_z$ 
(e.g. figure \ref{fig2}\emph{c}) whose effect 
is not considered in the layer model. 
A recent study \citep{15sreeni} shows that a horizontal magnetic field
of small axial lengthscale (more applicable to convection outside
the tangent cylinder than inside) shifts the viscous--magnetic
mode transition point from the classical value for a uniform field,
$\varLambda=O(E^{1/3})$ to a much higher value $O(1)$, 
without a drastic change in the critical
Rayleigh number. 
It is therefore possible that this transition point
is displaced to large $\varLambda_z$ for an axially varying $B_z$, 
allowing an intense, spatially varying
magnetic field to exist for $\overline{B^2} \sim 1$.
Linear magnetoconvection with field variation along the axial
coordinate $z$ (in addition to $x$, $y$ or both) 
brings additional complexities owing to
the presence of 
a  horizontal field component required to satisfy 
the divergence-free condition of
the mean field, and the presence of a
 mean flow 
arising from the Magnetic--Coriolis force balance in the vorticity
equation. Nevertheless, such a model is useful in predicting
the $z$-magnetic field intensity required to produce isolated
plumes within the tangent cylinder. 

\section{Concluding remarks}
\label{concl}
The new results that have come out of this study are 
summarized below in points (i)--(iv), 
together with what was known from earlier studies:
\begin{enumerate}
\item A comparison 
across Ekman numbers of the onset of an isolated
plume within the tangent cylinder in dynamo simulations
reveals that (a) the Rayleigh number for
plume onset increases
with decreasing $E$, and
(b) the plume width markedly decreases with 
decreasing $E$. These results bear the hallmark of
viscous-mode convection. In addition, the strong 
$B_z$--$u_z$ correlation suggests that the plume may seek out 
the location where the field is strongest.

Earlier studies (\S 1) proposed that the 
tangent cylinder plume is in the
large-scale magnetic mode, in which case
its onset Rayleigh number and width should be
independent of $E$.
However, these studies did not examine  
plume onset across Ekman numbers. 

\item  A laterally varying axial magnetic field
localizes convection in a 
rotating plane layer. The onset of convection 
takes the form of isolated plumes
in regions where the magnetic field is strong.
Of particular interest is the onset of
localized, small-scale convection 
(e.g. figure \ref{fig9}(\emph{c}),
for $\varLambda=0.04$ and $E=5 \times 10^{-5}$),
 in which case the critical Rayleigh number 
$Ra_c$ is not significantly
different from that for non-magnetic convection 
(figure \ref{fig6}(\emph{a}),
blue line).

Earlier onset models (see \S 1) did not consider the possibility 
of a laterally varying mean field locally exciting convection in
a rotating layer. These models predicted 
uniformly distributed convection either in the small-scale
viscous mode or in the large-scale magnetic mode, with the 
viscous--magnetic transition  
occurring at $\varLambda = O(E^{1/3})$.

\item The Rayleigh number for plume onset 
within the tangent cylinder agrees closely with the 
viscous-mode  Rayleigh number
in the plane layer 
magnetoconvection model (\S \ref{implications}). 
This result suggests that
the localized convection within the
tangent cylinder is in the viscous mode.

\item It follows from (iii) that the 
onset of an isolated plume within the tangent cylinder 
is approximately linear, even as nonlinear
dynamo action exists outside the tangent cylinder.

While it is already known that the onset of
pure (non-magnetic) convection inside the tangent cylinder
requires a Rayleigh number much higher than the critical
Rayleigh number outside it, our
study provides an analogous result for 
magnetic convection.

\end{enumerate}

Since the confinement of convection
occurs in both viscous and magnetic modes of onset
and the plume width increases at the mode cross-over
point (figures \ref{fig9}\emph{d} and \ref{fig10}\emph{b}),
it might appear that a strong magnetic field within
the tangent cylinder would give rise to a plume in the
magnetic mode. Notably, however, the plume width
does not increase with increasing $Ra$ (and $\varLambda_z$) 
in the dynamo regime of relatively strong rotation
($E= 5 \times 10^{-6}$; figure \ref{fig2}). This indicates
that the effect of rotation on the plume width 
prevails over the effect of the magnetic field, so that
the viscous--magnetic mode transition does not occur. 
It is hence reasonable to suppose that
the laterally varying field within 
the Earth's tangent cylinder would
strongly localize plumes in the small-scale
viscous mode, in turn producing non-axisymmetric polar vortices. 
The width of plumes is likely determined by the smallest scale
 that can  be supported against
magnetic diffusion in the core; 
it is plausible that this scale has magnetic
Reynolds number $Rm \sim 1$.

The nonlinear dynamo simulations in this study are
far from the low-$E$, low-$q$
regime thought to exist in the Earth's core. Simulations in which
magnetic diffusion is significantly higher than thermal
(and viscous) diffusion would help ascertain whether
the critical Rayleigh number for plume formation
progressively 
increases or tends to an asymptotic value as
$E$ is decreased.

The linear stability analysis
makes the simplifying assumption that the 
imposed field is invariant
along one of the horizontal
directions ($y$) that is chosen to be infinite in extent. 
The three-dimensional linear simulation
of the case in figure \ref{fig9}(\emph{c}) with the
field varying in both $x$ and $y$ shows
that the confinement in the $x$-direction is
merely replicated in the $y$-direction with no
change in the critical Rayleigh number
($Ra_c \approx 237.6$). Whereas decomposing the
perturbations as waves along $y$ involves
no loss of generality, it offers two distinct advantages:
The critical $y$-wavenumber 
($k_c$) readily confirms whether 
convection is viscously or magnetically controlled; and the
calculations are far less expensive than three-dimensional
onset simulations. Calculations for $E < 5 \times 10^{-7}$
are memory-intensive with the spectral
collocation method, but the evolution of pure spectral
methods may eventually overcome this limitation.

The confinement of rotating convection at small Elsasser number
does not imply that the mean magnetic field strength within
the Earth's tangent cylinder should be small. Rather, a field
strength of $\varLambda_z  \sim 10$ or higher is plausible
(table \ref{table0}). Consideration of
the axial inhomogeneity of the magnetic field would
likely displace the viscous--magnetic mode
cross-over point to much higher $\varLambda_z$,
which makes small-scale convection a reality for 
intense, spatially varying fields. Despite
the naturally occurring $z$-variation 
of the axial field inside the tangent cylinder,
the Rayleigh number for plume onset matches well 
with the approximately
constant Rayleigh number for viscous magnetoconvection. 
This indicates that the main
effect of the $z$-variation is to extend 
the viscous regime to higher Elsasser numbers.

Binod Sreenivasan thanks the Department of Science 
and Technology (Government 
of India) for the award of a SwarnaJayanti
 Fellowship. 
\vspace{0.1 in}
\appendix

\section{Matrices for linear magnetoconvection in two dimensions}
\label{app1}
The problem given by equations (\ref{mom2})--(\ref{current2})
is of the form $\bm{AX}=\lambda \bm{BX}$, where
\arraycolsep=5pt
\begin{eqnarray}
\bm{A} &=&
\begin{pmatrix}
\begin{array}{ccccc}
E D^4 & - I_x \otimes D_z & 0 & a_{14} & a_{15} \\
I_x \otimes D_z & E D^2 & 0 & a_{24} & a_{25} \\
I & 0 & q D^2 & 0 & 0 \\
a_{41} & f'(x)I (i k H) & 0 & D^2 & 0 \\
a_{51} & a_{52} & 0 & 0 & D^2 
\end{array}
\end{pmatrix}, \,\,
\bm{X}=
\begin{pmatrix}
\begin{array}{c}
u_z \\
\omega_z \\
\theta\\
b_z\\
j_z 
\end{array}
\end{pmatrix},\nonumber\\\nonumber\\
\arraycolsep=8pt
\bm{B} &=&
\begin{pmatrix}
	\begin{array}{ccccc} 
	0 & 0 & -q (D_x^2 \otimes I_z-k^2 I ) & 0 & 0 \\
	0 & 0 & 0 & 0 & 0 \\
	0 & 0 & 0 & 0 & 0 \\
	0 & 0 & 0 & 0 & 0 \\
	0 & 0 & 0 & 0 & 0 
	\end{array}
\end{pmatrix}, \,\,
\lambda=Ra. \label{abmatrix}
\end{eqnarray}

Here $I_x$ and $I_z$ are identity matrices of size $N_x \times N_x$ and $N_z \times N_z$ 
respectively ($N_x$ and $N_z$ being the number
of points in $x$ and $z$), so that $I= I_x \otimes I_z$ has size $(N_x \times N_z)^2$.

The differential operator matrices in (\ref{abmatrix}) are given by
\begin{align}
D^2 &= D_x^2 \otimes I_z + I_x \otimes D_z^2 - k^2 I, \\
D^4 &= D_x^4 \otimes I_z + k^4 I + I_x \otimes D_z^4
    - 2 k^2 D_x^2 \otimes I_z \nonumber\\
    &\quad - 2 k^2 I_x \otimes D_z^2 + 2 D_x^2 \otimes D_z^2, \\
H &= [D_x^2 \otimes I_z - k^2 I]^{-1}.
\end{align}
The abbreviated elements of matrix ${\bm A}$ are as follows:
\begin{align}
 a_{14} &= \Lambda \bigl[f(x)I D^2 (I_x \otimes D_z) 
   - 2 f''(x) I (D_x \otimes I_z) (H D_x \otimes D_z) \nonumber  \\
   &\qquad - f'(x)I (D_x^2 \otimes I_z) (H D_x \otimes D_z) + 
   f''(x)I (I_x \otimes D_z) \nonumber\\
   &\qquad + 2 f'(x)I( D_x \otimes D_z) - f'''(x)I (H D_x \otimes D_z) \nonumber\\ 
   &\qquad + k^2 f'(x)I (H D_x\otimes D_z) + f'(x)I (I_x \otimes D_z^2)
    (H D_x \otimes D_z)  \bigr],\\
  a_{15} &= \Lambda \bigl[-2 f''(x)I (D_x \otimes I_z) 
   (i k H) - f'(x)I (D_x^2 \otimes I_z) (i k H) \nonumber\\
   &\qquad -  f'''(x) I (i k H) + k^2 f'(x) I (i k H) + f'(x) I (I_x \otimes D_z^2) (i k H) \bigr],\\
   a_{24} &= \Lambda \bigl[- f'(x)I(I_x \otimes D_z) 
   (i k H I_x \otimes D_z)\bigr],\\
   a_{25} &= \Lambda \bigl[f(x)I (I_x \otimes D_z) + f'(x)I 
   (I_x \otimes D_z) (H D_x \otimes I_z)\bigr],\\
   a_{41} &= f(x) I (I_x \otimes D_z)+f'(x)I (H D_x \otimes D_z), \\
   a_{51} &= - f(x) I (I_x \otimes D_z)(ikH D_x \otimes D_z), \\
   a_{52} &= f(x)I (I_x \otimes D_z)+f'(x)I (I_x \otimes D_z)(H D_x \otimes I_z),
   \end{align} 

A standard approach is followed in the construction of the 
differentiation matrices
\citep{00tref,06huang}.
For $x=[0,L_x]$, the first order Fourier 
differentiation matrix for 
even $N_x$ is,
\begin{equation}
(D_x)_{ij}=  
\begin{dcases}
    0,& i=j,\\
    \frac{\pi}{L_x} (-1)^{i-j}\text{cot}\frac{(i-j)h}{2}, & i\neq j,
\end{dcases}
\end{equation}
and for odd $N_x$,
\begin{equation}
(D_x)_{ij}= 
\begin{dcases}
    0,& i=j,\\
    \frac{ \pi}{L_x} (-1)^{i-j}\text{csc}\frac{(i-j)h}{2}, & i\neq j,
\end{dcases}
\end{equation}
where $h=2 \pi/N_x$.

The transformed Gauss-Lobatto points for $z$ in the range $[0,1]$ are given by 
\begin{equation}
z_j=\frac{1}{2}\bigl(\text{cos}(j \pi /N_z)\bigr)+\frac{1}{2}, \ \ j=0, \ . \ . \ . \ . \ , N_z,
\end{equation}
and the first order Chebyshev differentiation matrix is given by
\begin{equation}
(D_z)_{ij}= 
\begin{dcases}
    \frac{2 N_z^2+1}{3}, & i=j=0, \\
    \frac{c_i}{c_j}\frac{(-1)^{i+j}}{z_i-z_j}, & i\neq j, \\
    \frac{-\text{cos}(j \pi /N_z)}{1-\text{cos}^2(j \pi /N_z)}, & 0<i=j<N_z, \\
     -\frac{2 N_z^2+1}{3}, & i=j=N_z, 
\end{dcases}
\text{where} \ c_i= 
\begin{dcases}
    2,& i=0,N_z.\\
    1, & \text{otherwise}.
\end{dcases}
\end{equation}

\section{Multiple unstable modes in a plane layer of finite aspect ratio}
\label{app2}

For stationary convection in a plane layer with
periodic $x$-boundaries spaced a length $L_x$ apart and stress-free $z$-boundaries
spaced unit distance apart, 
the axial velocity has the functional form
\begin{equation}
u_z (x,z) = A \sin (n \pi z) \exp (2 \pi i m/L_x).
\label{uzform}
\end{equation}
Following \cite{61chandra}, this solution is introduced into 
(\ref{mom2})--(\ref{temp2}) to give the characteristic equation
\begin{equation}
Ra=\frac{E}{a^2}\biggl[(n^2 \pi^2+a^2)^3 + \frac{n^2 \pi^2}{E^2}\biggr],
\end{equation}
where 
\begin{equation}
a^2 = \biggl(\frac{2m \pi}{L_x}\biggr)^2 + k^2. \nonumber
\end{equation}
Since $k \in\mathbb{R}^+$, for marginal state (critical) convection we obtain
\begin{equation}
m_c \leq \floor*{\frac{a_c L_x}{2 \pi}}. 
\label{modes}
\end{equation}
For $E= 1 \times 10^{-4}$, onset of convection occurs at 
$Ra_c=189.7$ and $a_c=28.02$. For $L_x=2$,
 $m_c$ can take 9 integer
values: $0, 1, 2, . \ . \ . \ , 8$. 
The corresponding critical $y$-wavenumbers are
\begin{equation}
k_c= 28.02, 27.84, 27.31, 26.39, 25.04, 23.20, 20.73,
 17.37, 12.39.\nonumber
\end{equation}
These 9 modes appear at the onset of convection 
(figure \ref{fig5}(\emph{a}), black line).  

In the presence of a uniform axial ($z$) magnetic field, the form
of the function in (\ref{uzform}) gives the following characteristic
equation \citep{61chandra}:
\begin{equation}\label{ra2}
Ra = \frac{E}{a^2}\frac{(n^2\pi^2 + a^2)\bigl(\bigl[(n^2\pi^2+a^2)^2 + 
(\Lambda/E) n^2\pi^2 \bigr]^2 + 
(1/E^2) n^2\pi^2 (n^2\pi^2+a^2)\bigr)}{\bigl[ (n^2\pi^2+a^2)^2+ (\Lambda/E) 
n^2\pi^2 \bigr]}.
\end{equation}
For $E=1 \times 10^{-4}$ and 
$\varLambda=0.5$, onset of magnetoconvection
occurs at $Ra_c=87.93$ and $a_c=3.35$. 
For $L_x=4$, (\ref{modes}) gives $m_c \leq 2$. The critical $y$-wavenumbers for
$m_c=0,1,2$ are therefore,
\begin{equation}
k_c= 3.35, 2.96, 1.16. \nonumber
\end{equation}
These 3 modes appear at the onset of magnetoconvection 
(figure \ref{fig5}(\emph{c}), blue line).

\bibliography{ref}

\bibliographystyle{jfm}

\end{document}